 \pdfoutput=1 
\documentclass{aa}  

\usepackage{txfonts}
\usepackage{color}
\usepackage{listings}
\usepackage{array}
\usepackage{subfigure}
\usepackage[dvipsnames]{xcolor}
\usepackage[colorlinks=true,linkcolor=MidnightBlue,citecolor=MidnightBlue,urlcolor=MidnightBlue]{hyperref}
\usepackage{comment}
\usepackage{graphicx}

\usepackage[capitalize,nameinlink]{cleveref}

\usepackage{soul}
\usepackage{threeparttable}
\usepackage{multirow}

\begin{document}

\title{The Solar System could have formed in a low-viscosity disc: \\ A dynamical study from giant planet migration to the Nice model }
\titlerunning{The Solar System could have formed in a low-viscosity disc.}

\author{P. Griveaud \inst{1}
\and A. Crida\inst{1}
\and A. C. Petit\inst{1}
\and E. Lega\inst{1}
\and A. Morbidelli\inst{1,2}}


\institute{Universit{\'e} C{\^o}te d'Azur, Observatoire de la C{\^o}te d'Azur, CNRS, Laboratoire Lagrange, France
\and
Collège de France, CNRS, PSL Univ., Sorbonne Univ., Paris, 75014, France}


    \abstract
  {In the context of low-viscosity protoplanetary discs (PPDs), the formation scenarios of the Solar System should be revisited. In particular, the Jupiter-Saturn pair has been shown to lock in the 2:1 mean motion resonance while migrating {generally} inwards, making the Grand Tack scenario impossible.}
  {We explore what resonant chains of multiple giant planets can form in a low-viscosity disc, and whether these configurations can evolve into forming the Solar System in the post gas disc phase.}
  {We used hydrodynamical simulations with the code FARGOCA to study the migration of the giant planets in a disc with viscosity parameter of $\alpha=10^{-4}$. 
  After a transition phase to a gas-less configuration, we studied the stability of the obtained resonant chains through their interactions with a disc of leftover planetesimals by performing N-body simulations using \texttt{rebound}.}
  {The gaps opened by giant planets are wider and deeper for lower viscosity, reducing the damping effect of the disc. {Thus, when planets enter a resonance, the resonant angle remains closer to circulation, making the chain weaker}. Exploring numerous configurations, we found five stable resonant chains of four or five planets. 
  In a thin (cold) PPD, the four giant planets revert their migration and migrate outwards.
  After {disc dispersal, under the influence of a belt of planetesimals,} some resonant chains undergo an instability {phase} while others {migrate smoothly over a billion years.}
  For three of our resonant chains, about $\sim 1\%$ of the final configurations pass the four criteria to fit the Solar System. 
  {The most} successful runs {are obtained} for systems formed in a cold PPD with a massive planetesimal disc.} 
  {This work provides a fully consistent study of the dynamical history of the Solar System's giant planets, from the protoplanetary disc phase up to the giant planet instability.
  Although building resonant configurations is difficult in low-viscosity discs, we find {it} possible to reproduce the Solar System from a cold, low-viscosity protoplanetary disc.}

       \maketitle
      



\section{Introduction}

Our Solar System is clearly the most studied planetary system in the field. This makes it  an excellent laboratory to test the theories of planet and planetary systems formation; however,   its extremely well-constrained features make it difficult to fit our models. Since the discovery of the wide variety of exoplanetary systems, we can now qualify our Solar System as uncommon. Hence, general planet formation theories should make the Solar System a possible outcome, but not the most likely one.

A variety of models focusing on different periods of the Solar System's lifetime have been proposed to explain the plethora of features observed today.
On the one hand, during the protoplanetary disc (PPD) phase, the Grand Tack Scenario \citep{walshLowMassMars2011} uses the  migration of Jupiter and Saturn to possibly explain the low mass of Mars with respect to the Earth and the variation in the composition of the asteroid belt. In this model, Jupiter would have migrated inwards until reaching $\sim 1.5$ au, the point at which Saturn would get locked in the 3:2 mean motion resonance (MMR) with Jupiter triggering a reversal in the direction of migration due to the \cite{massetReversingTypeII2001} mechanism. {However, the Grand Tack model is losing appeal as alternative mechanisms have been found to explain the low mass of Mars and the depletion of the asteroid belt \citep[e.g.][]{drazkowskaCloseinPlanetesimalFormation2016, clementMarsGrowthStunted2018, clementEarlyInstabilityScenario2019, nesvornyRoleEarlyGiantplanet2021, morbidelliContemporaryFormationEarly2022}. However, blocking the inward migration of Jupiter remains a requirement to explain why it did not acquire a semi-major axis of 1-2 au, typical of most observed Jovian-mass planets around other stars.}

On the other hand, once the protoplanetary disc dissipates the giant planets are in a compact resonant configuration, while today's Solar System is in a very different situation.
Thus, another model is required to explain the final stage of formation of the Solar System: the Nice Model. It is an appealing scenario that explains how the planets were brought to their current positions through a violent instability. This instability was provoked by the gravitational interactions between the giant planets and a disc of leftover planetesimals located at the edge of the Solar System. This model is able to reproduce {simultaneously} different constraints in today's System, starting with the orbital configuration of the giant planets as well as Jupiter's trojan asteroids, or the high eccentricities and inclinations of the asteroids in the asteroid belt and the leftover Kuiper belt \citep[see][for a review]{nesvornyDynamicalEvolutionEarly2018}. 

The choice of the initial planetary configuration and the characteristics of the instability have evolved through the different studies of the Nice Model. In the original Nice Model \citep[hereafter  ONM;][]{tsiganisOriginOrbitalArchitecture2005,morbidelliChaoticCaptureJupiter2005,gomesOriginCataclysmicLate2005} the focus was set on the trigger of this instability: the crossing of the 2:1 MMR from the inside out between Jupiter and Saturn.
As the planets exchange angular momentum with the planetesimals, they migrate outwards for Saturn, Uranus and Neptune and inwards for Jupiter, which ejects the planetesimals out of the Solar System. As they undergo divergent migration Jupiter and Saturn approach the 2:1 MMR, this excites their eccentricities and the System eventually goes unstable. The remaining planetesimals   help damp the planetary orbits back to the stable configuration we know today. The ONM explains the non-zero eccentricities of Jupiter and Saturn. It is also consistent with the population of irregular satellites, Jupiter's trojans, the distribution of asteroid in the main belt and finally with the scenario of the late heavy bombardment (LHB). 

The LHB at the time of the first Nice Model studies was one of the main constraints for the giant instability to happen late in the Solar System history (at about $\sim 700$ Myr). However, in the ONM the timing of the instability was highly dependent on the initial position of the disc of planetesimals.  Another weakness of the ONM concerns the arbitrary initial configuration of the planets. 
As a result of planet migration during the gas disc phase, the planets most likely ended in a resonant configuration by the time the disc dispersed, which was not accounted for in the ONM. At the time, migration studies found Jupiter and Saturn most likely locked in a 3:2 MMR \citep{massetReversingTypeII2001,morbidelliDynamicsJupiterSaturn2007}. Hence, \cite{morbidelliDynamicsGiantPlanets2007} extended the Jupiter-Saturn study by adding Uranus and Neptune. They found six possible resonant chains, two of them  stable for more than 100 Myr, potentially destabilised later by a disc of planetesimals. This led to a renewal of the Nice Model, later called the Nice Model 2 \citep{levisonLATEORBITALINSTABILITIES2011}, in which the initial configuration of the planets was taken from hydrodynamical simulations accounting for planet migration in the disc phase. 
{In this new version of the model, the authors found that the instability could occur late for a broader range of initial conditions, provided that initially the planetesimal disc is far enough from Neptune to avoid any planet--planetesimal scattering.}

{However, since then the chronology of lunar bombardment has been revised. It is a fact that the lunar bombardment was heavy 3.9 Gyr ago (i.e. 600My after lunar formation). However, while in the aftermath of the Apollo mission it was thought that such a heavy bombardment was the result of a sudden surge in the collision rate, following a quiescent era \citep{teraIsotopicEvidenceTerminal1974,ryderCatastrophicEventsMass2000}, new analyses of impact ages on the individual minerals of the lunar samples and in lunar meteorites show that there was no evident paucity of impacts prior to that time \citep[see][]{zellnerCataclysmNoMore2017}. A monotonic decline in the bombardment rate has been shown to be consistent with dynamical simulations, the lunar crater record, and the lunar late veneer, provided that the Moon lost trace of the craters and of the highly siderophile elements delivered before 4.35 Gyr ago, which would be the case if the lunar mantle had remained partially molten up to that time \citep{morbidelliTimelineLunarBombardment2018,zhuReconstructingLateaccretionHistory2019,nesvornyEarlyBombardmentMoon2023}. Thus, there is no need to invoke a late giant planet instability to explain the bombardment rate of the Moon 3.9 Gyr ago.}

The most extensive study of the Nice Model instability came from \cite{nesvornySTATISTICALSTUDYEARLY2012}. In that work, the authors ran a statistical study of planetary instability with a wide parameter space including initial positions of planets, position, and mass of the planetesimal disc as well as the number of planets initially present in the system. 
Today's orbital excitation of Jupiter and the presence of its irregular satellites and Trojan asteroids show evidence that the gas giant must have experienced an encounter with another planet. As a consequence that planet is easily ejected out of the system. Hence, if only four planets existed at the beginning the system, it fails to reproduce today's Solar System. 
\cite{nesvornySTATISTICALSTUDYEARLY2012} thus explored the possibility of having a third and even a fourth ice giant, which would eventually be ejected by Jupiter. 
In their work, though, the initial positions of the planets are set using a migration prescription mimicking the gas disc phase. While this allows for a cheaper and thus more extensive exploration of resonant chain building, it does not reproduce all the complexity of planet-disc interactions as a hydrodynamical simulation would.

Although we do not know the exact initial configuration of the giants planets at the end of the disc phase, it is important to consider realistic migration history as a starting point of the Nice Model. 
The physics of protoplanetary discs can have quite some consequences on the building of resonant planetary systems.
At the time of all the previously cited studies, protoplanetary discs were believed to be highly viscous. However, it is now believed that discs might not be so turbulent {(see \cite{turnerTransportAccretionPlanetForming2014} or \cite{lesurHydroMagnetohydroDustGas2022} for reviews on the theoretical arguments, and \cite{pinteDustGasDisk2016, dullemondDiskSubstructuresHigh2018,sellekDustyOriginCorrelation2020, villenaveHighlySettledDisk2022} for observational constraints)}.
{In addition to the   results of \cite{pierensOutwardMigrationJupiter2014}, we showed recently in \cite{griveaudMigrationPairsGiant2023}} that the migration of pairs of planets in low-viscosity discs is significantly different than in a high-viscosity case: Jupiter and Saturn are locked in a 2:1 MMR, instead of a 3:2. {However, unlike \cite{pierensOutwardMigrationJupiter2014}, in our study we do not find any outward migration of the pair in this configuration. We showed that this result is robust with respect to modifications in disc mass, aspect-ratio, and initial positions and masses of the outer planet.
As was mentioned in our previous paper, the two studies differ in the treatment of the energy equation and in the surface density profile; these differences could perhaps explain the difference in migration direction of the pair.}
{Nevertheless, from our previous study we could conclude} that the Grand Tack scenario mentioned above, would not be possible if the Solar System's protoplanetary disc had a low-viscosity. 
However, as said before, the Grand Tack is no longer necessary, thus opening the possibility of a low-viscosity proto-solar nebula to be considered, although it would be important to find a mechanism to block Jupiter’s inward migration.
Therefore, in this paper, we are interested in extending the \cite{griveaudMigrationPairsGiant2023} results to the Nice Model.

For reference, most of the previously cited studies of the Nice Model had Jupiter and Saturn in a 3:2 configuration.   Some studies sparsely considered a few resonant chains with Jupiter and Saturn in a 2:1 MMR, mostly for academic purposes \citep{nesvornySTATISTICALSTUDYEARLY2012, batyginEarlyDynamicalEvolution2010, thommesMeanMotionResonances2008, pierensOutwardMigrationJupiter2014}. More recently, \cite{clementBornEccentricConstraints2021,clementBornExtraeccentricBroad2021} studied extensively instabilities with Jupiter and Saturn in a 2:1 resonance {modulating the eccentricities of the two planets, with values ranging from 0.025 to 0.25}.
However, all these studies built resonant chains regardless of the physics of protoplanetary discs, mimicking planetary migration in a simplified method. In contrast, \cite{cridaMinimumMassSolar2009} studied with hydrodynamical simulations the migration of the four Solar System giant planets in viscous discs of various density profiles, but did not address the post-disc phase. To our knowledge,
only \cite{morbidelliDynamicsGiantPlanets2007} united both phases in a single study, thus creating a bridge between planet migration and the giant planets' instability. 
Similar to their work, the aim of this paper is to run a self-consistent study of the orbital evolution of the Solar System's giant planets
from the protoplanetary disc phase to today's configuration, in the context of low-viscosity discs. We want to first study how resonant configurations are formed in such discs and how low-viscosity influences the building of these systems. Then, we use the outputs of hydrodynamical simulations as initial conditions to N-body simulations of the Nice Model and, while \cite{morbidelliDynamicsGiantPlanets2007} only showed a few proof-of-concept N-body simulations of the global instability, we  assess the success rate of the reproduction of the Solar System.

The paper is divided in two main sections. \Cref{sec:hydro} explores the phase of planetary migration within the protoplanetary disc. We first describe how we ran the hydrodynamical simulations in \cref{sec:hydro_methods}. Then in \cref{sec:hydro_results}, we present the results of the migration of the four or five planets for two different disc scale heights. In order to transition towards the N-body simulations, we mimic the gas disc dissipation and discuss the consequences on the planetary system in \cref{sec:evap}.
\Cref{sec:hydro_conclusion} presents a `half-way' conclusion on the formation of resonant chains in low-viscosity discs. 
The paper continues with \Cref{sec:Nbody}, which studies the giant planet instability phase from the previously built resonant chains and after the gas disc dispersed. We present our methods in \cref{sec:Nbody_methods} and recall how to assess the success of a simulation based on the criteria of \cite{nesvornySTATISTICALSTUDYEARLY2012} in \cref{sec:criterions}. The results are presented in \cref{sec:Nbody_results}. Finally, we discuss and conclude our complete study in \cref{sec:conclusion}.

\section{Protoplanetary disc phase: Hydrodynamical study} \label{sec:hydro}

In the hydrodynamical study section of this paper, we re-used the simulations of \cite{griveaudMigrationPairsGiant2023}, which covered extensively the migration of Jupiter and Saturn {in discs with different properties}, adding ice giants in the outer regions of the disc.
We recall in the following {sub}section some methods used in our previous paper (for more details see Sect. 2 of \cite{griveaudMigrationPairsGiant2023}). {Our new results are introduced in the next subsection.

\subsection{Methods} \label{sec:hydro_methods}

%
We ran 2D global hydrodynamical simulations with the grid-based code FARGOCA.\footnote{{A recently re-factorised version of the code that can be found at: https://gitlab.oca.eu/DISC/fargOCA}}
We set the \cite{shakuraBlackHolesBinary1973} viscosity parameter to {be uniform and constant as} $\alpha=10^{-4}$.  

The two-dimensional system is described in polar coordinates $(r,\phi)$, centred on the star, where $r$ is the radial coordinate and $\phi$ the azimuthal coordinate. 
The disc's aspect ratio follows the radial profile $h=h_0(r/r_0)^{2/7}$, where $r_0$ is the unit length, $h_0=0.05$ in what we further on call the {nominal} simulation, and $h_0=0.035$ in the {cold} simulations.
The disc density is given by $\Sigma = \Sigma_0 (r/r_0)^{-1/2}$ where $\Sigma_0 = 6.76\cdot 10^{-4}$ in code units for the {reference} case, which makes $222 \,\text{g cm}^{-2}$ for $r_0 = 5.2$ au and a Sun mass star. 

We considered the four giant planets of our Solar System; Jupiter with mass $M_J/M_\star = q_J = 10^{-3}$, Saturn $ q_S = 2.9 \cdot 10^{-4}$, Uranus $q_U = 4.4 \cdot 10^{-5}$ and Neptune $q_N = 5.1 \cdot 10^{-5}$. In some cases, we also added a third ice giant in our system of Uranus mass. 
We find that systems are easily unstable with three ice giants, so we also considered the possibility of a less massive planet, taking $q_{P9} = 1.9 \cdot 10^{-5}$. This is motivated by the Planet Nine hypothesis \citep{batyginEvidenceDistantGiant2016, batyginPlanetNineHypothesis2019} according to which there exists a ninth planet in the Solar System located on a highly eccentric orbit of mass estimated around six Earth masses \citep{brownOrbitPlanetNine2021}. This planet could have reached such orbit after a close encounter with one of the gas giants and through interactions with passing stars \citep{batyginPlanetNineHypothesis2019}. 
In the case of Jupiter and Saturn, the planets mass was increased over {800 initial} orbital periods in order to perturb smoothly the gas disc \citep[see][for further details]{griveaudMigrationPairsGiant2023}. However, for the ice giants or planet Nine this was not necessary and thus the planets were added with their final mass into the simulation.

The radial domain extends in the range $r\in [0.2,9]r_0$, which corresponds to $r\in [1.04,46.8]$ {au} for $r_0=5.2$ {au}. The {grid} resolution is such that $dr/r = d\theta = 0.0067$, corresponding to $N_{\text{rad}}=568$ logarithmically spaced radial cells and $N_{\text{sec}}=940$ azimuthal cells. {This resolution gives at least 5 grid cells per scale height at $r_0$ and per half width of the horseshoe region for the four giant planets.} The boundary conditions used in the radial direction follow the prescription of the evanescent boundary condition \citep{deval-borroComparativeStudyDiscplanet2006}.

\subsection{Building resonance chains} \label{sec:hydro_results}

In this section, the aim is to find a maximum of resonance chains in a low-viscosity disc using four and five giant planets. Two different disc scale heights were considered in this exploration, since in \cite{griveaudMigrationPairsGiant2023}, we found Jupiter and Saturn in different resonant configuration depending on the scale height. Other disc parameters did not vary much the result of the migration of the pair.

\subsubsection{Nominal disc - $h=0.05$}\label{sec:nominal}

\begin{figure*}
    \centering
    \includegraphics[width=\textwidth]{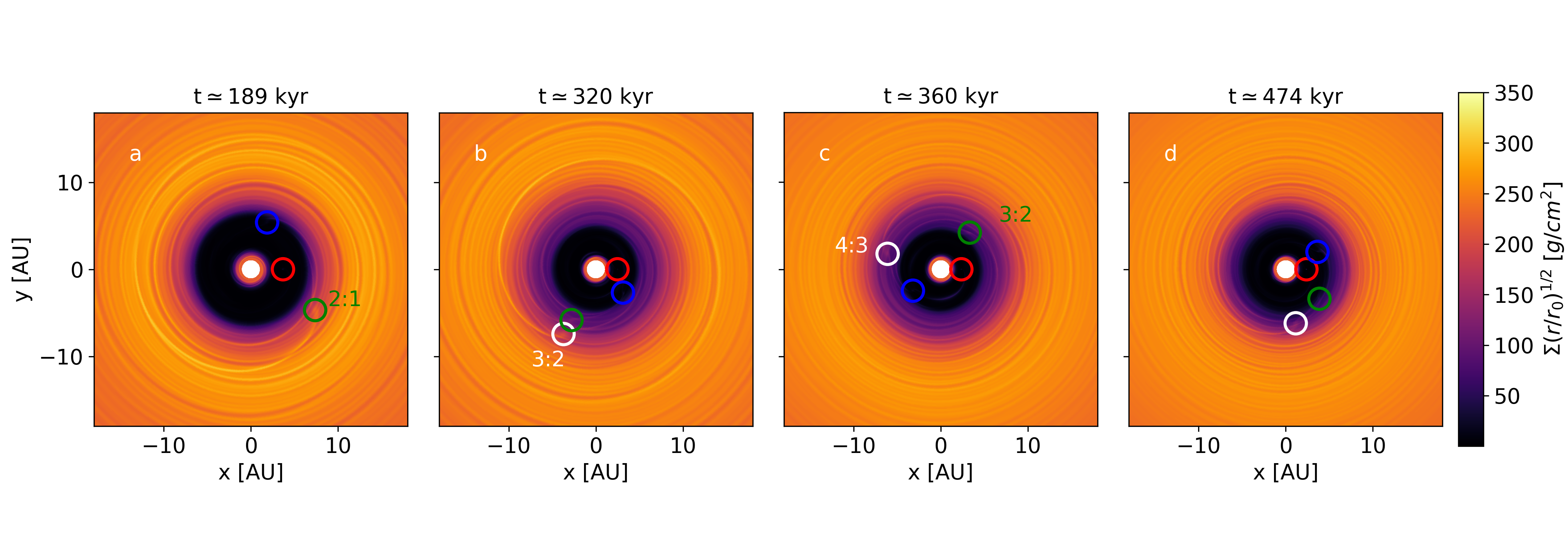}
    \caption{Time snapshots of the flattened surface density of the disc in the nominal simulation, $N4$.  The empty circles give the positions of Jupiter \textit{(red)}, Saturn \textit{(blue)}, Uranus \textit{(green),} and Neptune \textit{(white)}. Panel a shows the entry of Uranus in the 2:1 MMR with Saturn. Panel b shows the moment when Neptune has just been locked in the 3:2 resonance with Uranus. In panel c, the four planets enter into their final resonant configurations (2:1, 3:2, 4:3). Finally, panel d shows the stable system at the beginning of the mock disc dispersal phase (see \cref{sec:evap} for more details). These moments are snapshots in the evolution of the planetary orbits that can be found in \cref{fig:Nom}.}
    \label{fig:Hydro_SigmaMap}
\end{figure*}
\begin{figure}
    \centering
    \includegraphics[width=\columnwidth]{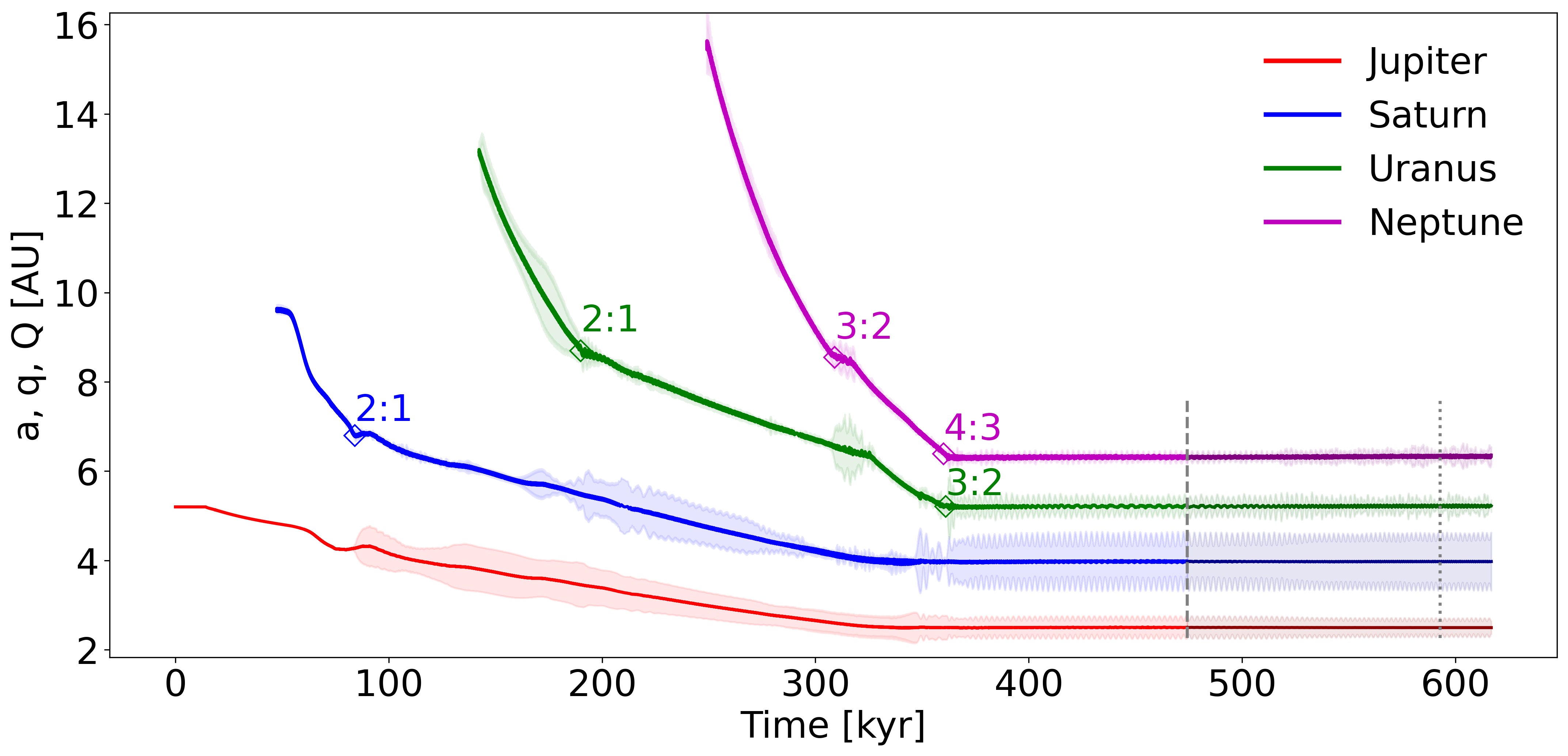}
    \caption{Evolution of the orbital parameters  of the four planets in the nominal simulation $N4$. The shaded areas mark the positions of the peri- and apo-centre of the planets, $q=a(1-e)$ and $Q=a(1+e)$,  respectively.
    The beginning of the plot up to about $150\,000$ years shows the nominal simulation $N$ of \cite{griveaudMigrationPairsGiant2023}. 
    Uranus and Neptune are added consecutively in the simulation and the final resonance chain is (2:1, 3:2, 4:3). Here Neptune's introduction time was $T_{N,0} = 250\,000$ years (see \cref{fig:NomBis}). The dashed and dotted vertical lines indicate the time at which the mock disc dispersal is respectively started ($t_0$) and ended ($t_f$); see section \ref{sec:evap} for more details.}
    \label{fig:Nom}
\end{figure}

As a follow-up to \cite{griveaudMigrationPairsGiant2023}, we added ice giants in the nominal simulation $N$ of the previous paper, so in a disc with aspect ratio $h=0.05$. 

A Uranus mass planet was added in the system at a radial distance of $2.5r_0$ from the star, just above twice the semi-major axis of Saturn. The planet migrates inwards and gets locked in the 2:1 MMR with Saturn at about $190$ kyr, the surface density of the disc at this moment is shown in panel a of \cref{fig:Hydro_SigmaMap}. The system with the three planets continues to migrate inwards, with a migration speed comparable to that of Jupiter and Saturn alone.
At about $250$ kyr, a fourth planet with Neptune's mass was added in the system at a distance of $3r_0$. As it migrates towards the system, it approaches the 3:2 MMR with Uranus and gets shortly locked in this resonance. However, this configuration excites both planets' eccentricities which eventually leads to Uranus being ejected out of both resonances with Neptune and Saturn. Panel b in \cref{fig:Hydro_SigmaMap}, shows the surface density of the disc precisely at the time when Neptune exits the 3:2 resonance with Uranus at about $320$ kyr. \\
The two ice giants then continue to migrate inwards until Uranus gets locked in the 3:2 MMR with Saturn. {At the same time, Neptune locks in the 3:4 MMR with Uranus, coincidentally the 2:1 MMR with Saturn.} 
The moment of this capture is shown in panel c of \cref{fig:Hydro_SigmaMap}.
Finally, the system remains stable in the resonant chain (2:1,3:2,4:3) for several hundred thousand years. The four planets sit in a common wide gap in the disc as shown in panel d of \cref{fig:Hydro_SigmaMap}. 
In such configuration, the eccentricities of the two gas giants are higher than the ice giants.  Jupiter and Saturn end up with eccentricities of about $0.09$ and $0.13$ respectively, while Uranus and Neptune's are of the order of $0.03$ and $0.01$. The resonances tend to excite the planets eccentricities while the gas disc damps them. This explains why the ice giants, being closer to edge of the gap, have lower eccentricities than the two more massive planets which are located in the deep and wide part of the gap. This also contrasts with the case of viscous discs where for instance in \cite{morbidelliDynamicsGiantPlanets2007} the most eccentric planet was Uranus with $e\approx 0.05$. 
Lastly, we note that the migration of the four planets stops when the ice giants lock in their final resonances. This {stopping} is due to the inner edge of the gap reaching the boundary of the grid and therefore cannot be interpreted as an effect of the ice giants. In fact, in a similar simulation without Uranus and Neptune, Jupiter and Saturn also stop migrating near the edge, while if we move the inner edge of the grid, they keep migrating inwards. {Furthermore, since the artificial stopping of Jupiter and Saturn did not force Uranus to cross the 3:2 MMR (nor Neptune the 4:3), we are confident that this resonant chain is robust. Therefore, this does not impact the purpose of this work. The next step is to mimic} the gas disc's dispersal (see \cref{sec:evap}).
We name this simulation $N4$ and show the evolution of the planetary system in Fig.~\ref{fig:Nom}. \\

The previous case, involving a complex dynamics with resonance breaking and capture, may have a strong sensitivity to initial conditions. Exploring the stability of such systems, we tried introducing Neptune at a different time. This is shown in \cref{fig:NomBis}. In this other simulation {(hereafter named $N_{\rm bis}4$)}, Neptune is introduced in the system at time $T_{N,0} = 238\,000$ years (while at time $T_{N,0} = 250\,000$ years in $N4$).
As previously seen, Neptune kicks Uranus out of the 2:1 MMR with Saturn. However, now, this triggers a succession of close encounters between Uranus and the two gas giants, which eventually leads to the ejection of Uranus out of the system (see Fig.~\ref{fig:NomBis} at $355\cdot 10^3$ years).
Despite this short instability moment, Neptune is captured in the 2:1 MMR with Saturn and the system of now three planets remains stable. Adding a new ice giant of Uranus mass, leads to a stable system of four planets in the following resonant chain (2:1, 2:1, 4:3). Having Uranus and Neptune inverted radially in the disc is not a problem as previous studies of the Nice Model show that in a significant fraction of simulations the two ice giants exchange their radial order. 

\begin{figure}
    \centering
    \includegraphics[width=\columnwidth]{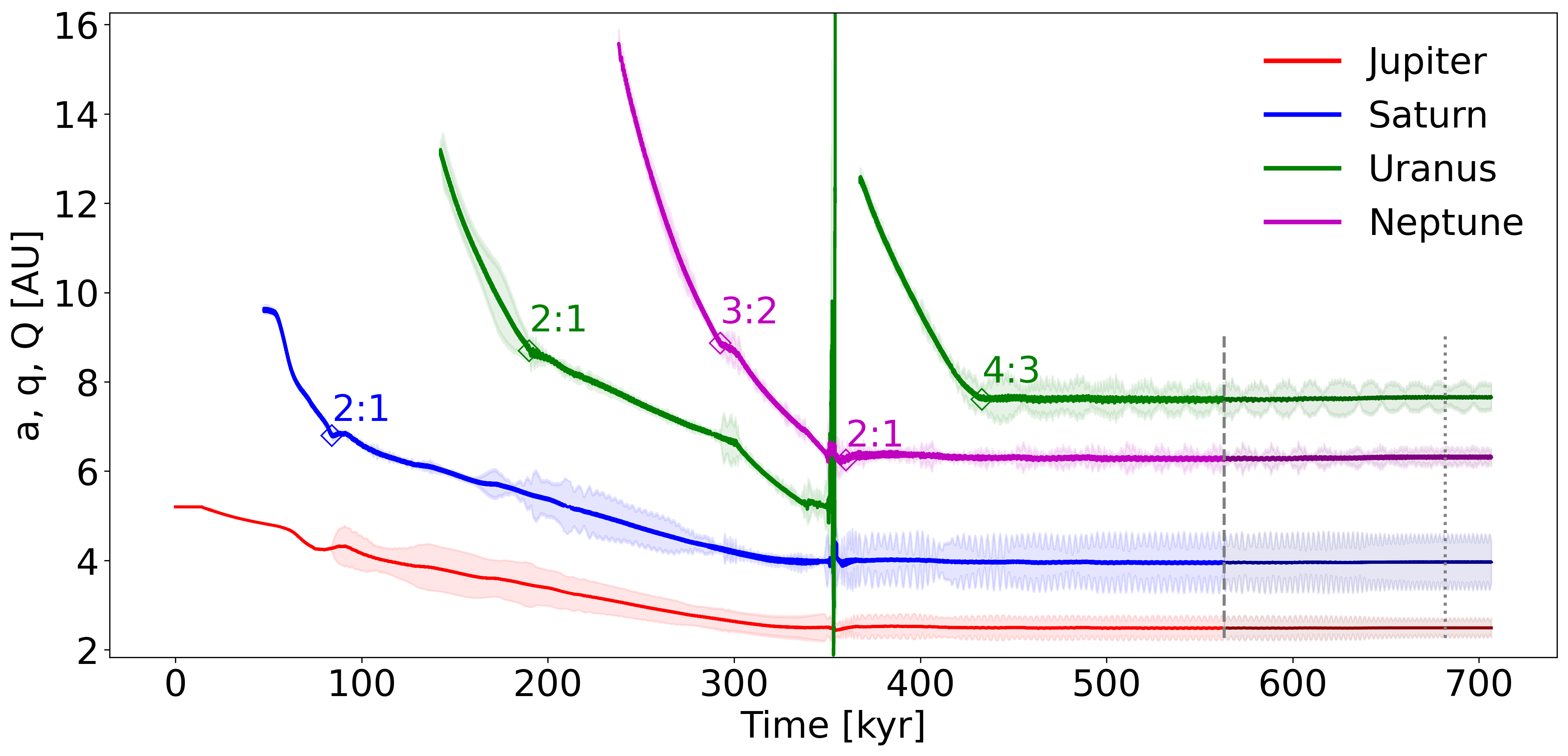}
    \caption{Same as \cref{fig:Nom}, but with Neptune's introduction time $T_{N,0} = 238\,000$ years, namely simulation $N_{\rm bis}4$. Here Uranus is ejected from the system at $355\,000$ years. We add another Uranus-mass ice giant later in the simulation and obtain the final resonance chain (2:1, 2:1, 4:3).}
    \label{fig:NomBis}
\end{figure}

\begin{figure}
    \centering
    \includegraphics[width=\columnwidth]{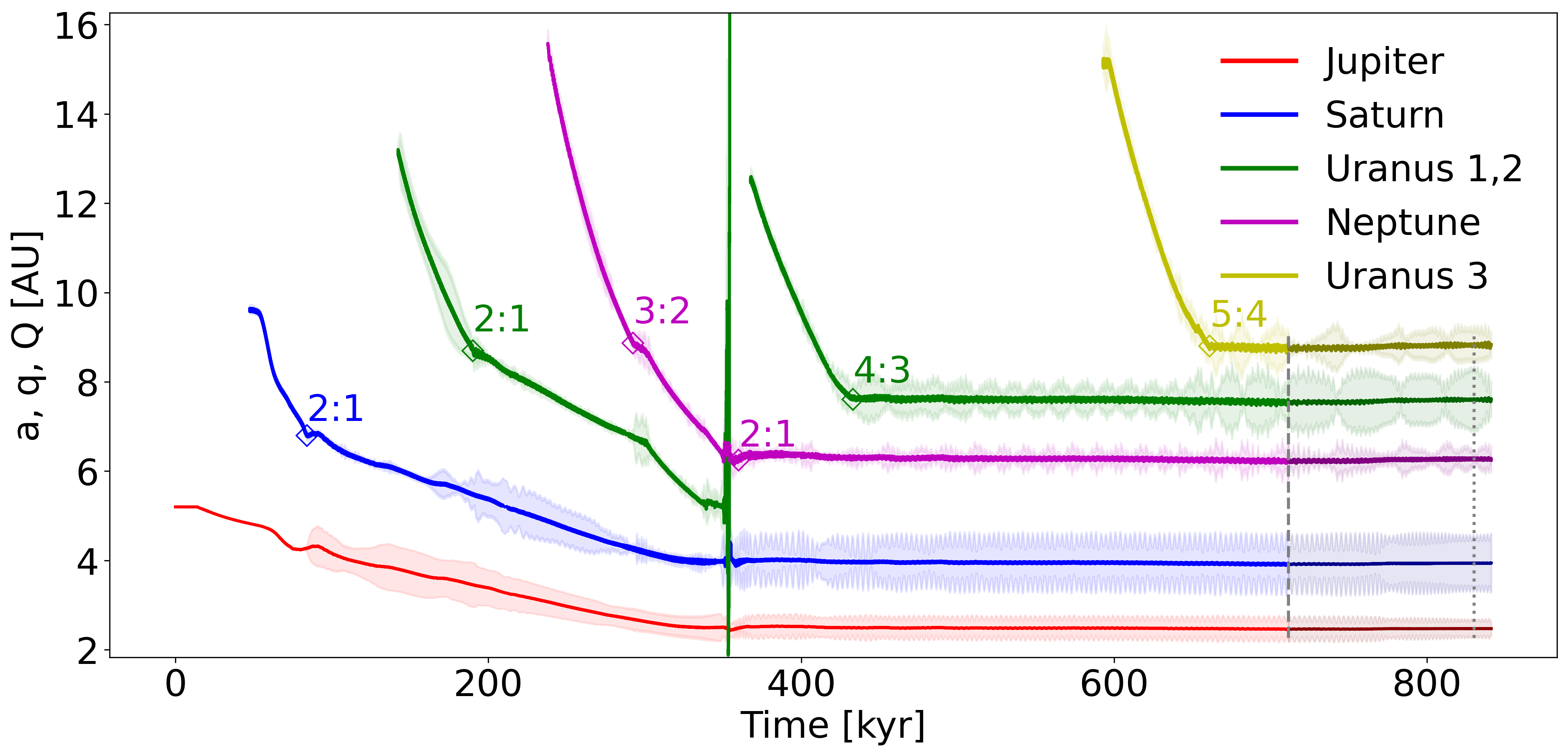}
    \caption{Same as \cref{fig:NomBis}, but  for simulation $N_{\rm bis}5$, in which an additional ice giant of Uranus mass was added in the simulation and  the final resonance chain (2:1, 2:1, 4:3, 5:4) was obtained. This simulation is stable during the gas disc phase; however, it becomes unstable during the mock evaporation phase (see \cref{sec:evap} for more details).}
    \label{fig:NomBis_5Giants}
\end{figure}

With two stable resonant chains of four planets, we explored the building of five-planets systems by adding another planet in each of these simulations. These trials are reported in \cref{tab:Hydro}.
In $N4$, the system went unstable (i.e. leading to the ejection of at least two planets) in two different attempts with a third ice giant, varying its introduction time. Assessing whether the stability could be achieved with a smaller mass planet, in a third attempt we added a Planet Nine like candidate, but this run did not succeed either. The resonant chain in simulation $N4$ is quite compact and therefore easily destabilised by the arrival of a fifth planet. 
In $N_{bis}4$, one case with a third ice giant went unstable, but changing its introduction time finally led to a stable configuration with five planets. This simulation, named $N_{bis}5$, is shown in \cref{fig:NomBis_5Giants} and the final resonant configuration is (2:1, 2:1, 4:3, 5:4).  


Lastly, we also explored the possibility of the fifth planet being Planet Nine and initially located in between the gas giants and the ice giants. So starting from simulation $N$ of \cite{griveaudMigrationPairsGiant2023}, we added in that order Planet Nine, Uranus then Neptune into the simulation. However, this was not successful in forming a stable configuration. 

In this sub-section, the exploration was accomplished changing the introduction of the planets as well as the number of planets in the configuration. However, we have not modified any physical disc parameters at this point. 

\subsubsection{Cold disc - $h=0.035$}\label{sec:cold}

In \cite{griveaudMigrationPairsGiant2023}, we ran a wide parameter exploration to study the resonant configuration of Jupiter and Saturn. We found that in most cases they are locked in the 2:1 MMR and their migration behaviour remains similar throughout the exploration, except in a disc with aspect ratio of $h_0=0.035$. 
{In the low-viscosity context, this value is consistent with values predicted for passive discs (that is heated by stellar radiation only) by \cite{chiangFormingPlanetesimalsSolar2010} or slightly above those predicted by \cite{savvidouInfluenceGrainGrowth2020}.} 
In such thinner (and thus colder) disc, Jupiter and Saturn are locked in the 5:2 MMR. 
Although this result was unexpected, we showed {in \cite{griveaudMigrationPairsGiant2023}} that this resonant configuration was a consequence of a density maximum located outside of Jupiter's gap which acted as a planet trap for the growing Saturn. Hence, we explored this case with more planets. 
A Uranus mass planet was added in simulation $C$ of \cite{griveaudMigrationPairsGiant2023}. We found that as it migrates towards Saturn the two planets lock in the 2:1 resonance. This pushes Saturn out of the 5:2 with Jupiter and inwards until it reaches the 2:1 MMR. This shows that the 5:2 MMR between Jupiter and Saturn is not a robust resonant configuration. Nevertheless, the system reaches a stable resonant configuration. 
At this point the migration of the three planets is similar to the case where Jupiter and Saturn were already in a 2:1 MMR in a cold disc (simulation $C_3$ in \cite{griveaudMigrationPairsGiant2023}). So for simplicity, in the following, we only consider the case of Jupiter and Saturn in the 2:1 in the cold disc case. 

In \Cref{fig:Cold_4Giants} we show the evolution of the orbital elements of the four planets in a cold disc, {namely simulation $C4$}. Uranus gets locked in the 3:2 MMR with Saturn. 
After being introduced, Neptune migrates inwards and starts opening a gap thus pushing an excess mass towards Uranus. 
This density bump then pulls Uranus out of the resonance towards Neptune. This is shown in the background of \cref{fig:Cold_4Giants}, where the azimuthally averaged surface density is plotted as a function of time.
Both planets eventually continue their migration inwards in a common gap until the system reaches the final resonant chain (2:1, 3:2, 5:4). The four giant planets now lie in a common wide and deep gap, promoting outward migration.

Similar as in the nominal disc, when adding a fifth ice giant in simulation $C4$, we find again that the system is highly unstable, despite multiple attempts varying the introduction time of the fifth planet. These attempts are reported in \cref{tab:Hydro}. Instead, we were able to find a stable configuration with a Planet Nine candidate, this simulation $C5$ is shown in \cref{fig:Cold_5Giants}.
The low mass planet is trapped at a density maximum located at the edge of the gap \citep{massetDiskSurfaceDensity2006}, which coincidentally is near the 4:3 MMR with Neptune. However, during this phase, the resonant angles corresponding to the 4:3 MMR do not librate. During the gas dispersal phase (see \cref{sec:evap}) P9 definitely enters the 4:3 resonance and thus the five planets end up forming the following resonant chain: (2:1, 3:2, 5:4, 4:3).

\begin{figure}
    \centering
    \includegraphics[width=\columnwidth]{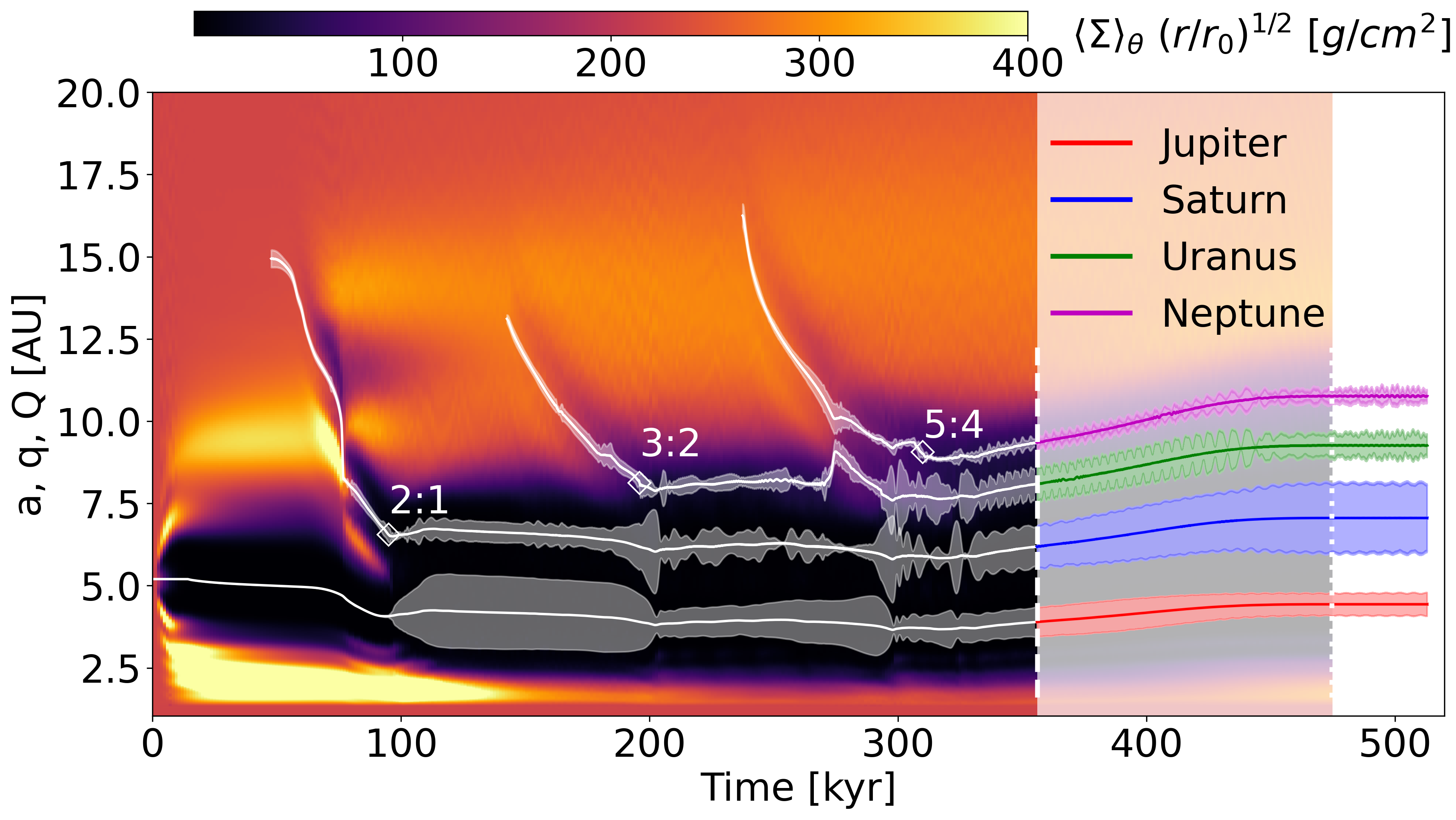}
    \caption[ ]{Similar to \cref{fig:Nom}, this figure shows the orbital parameters of the four giant planets evolving in a colder disc, so lower aspect ratio, simulation $C4$.\protect\footnotemark\
    Additionally, the background of the figure shows the azimuthaly averaged surface density profile of the disc as a function of time throughout the simulation. For the sake of clarity, the left half of the plot shows all the planets' orbital parameters in white. In the second half of the plot, from about 350 kyr, the phase of the gas disc's dispersal is initiated. The vertical dashed and dotted lines mark the start $t_0$ and end $t_f$ of this phase (see section \cref{sec:evap} for more details). The system is stable in the resonance chain (2:1, 3:2, 5:4).}
    \label{fig:Cold_4Giants}
\end{figure}
\footnotetext{The first $150$ kyr of this simulation correspond to simulation $C_3$ of \cite{griveaudMigrationPairsGiant2023}.}

\begin{figure}
    \centering
    \includegraphics[width=\columnwidth]{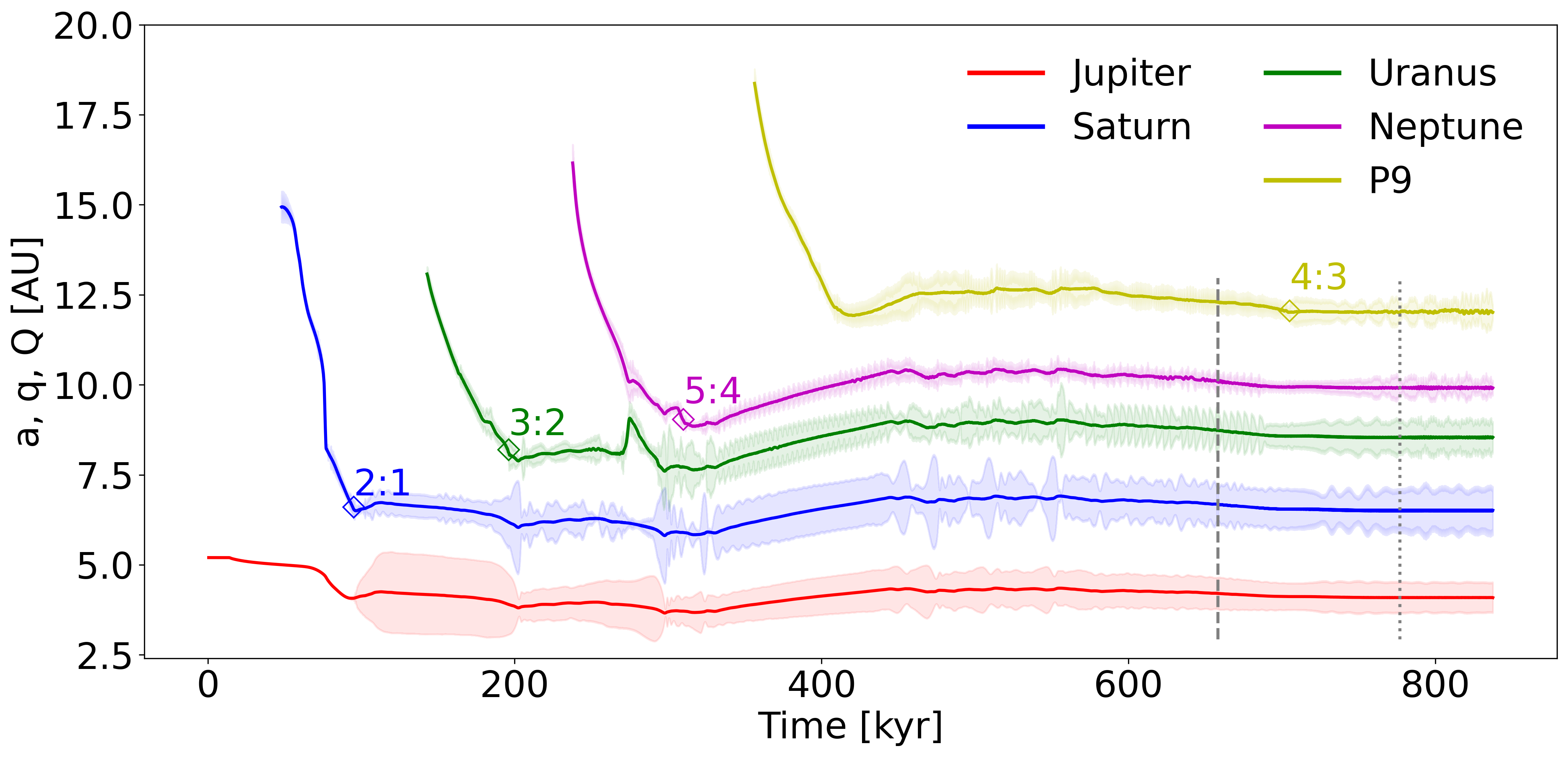}
    \caption{Similar to \cref{fig:Cold_4Giants}, but for simulation $C5$, so with an additional planet of mass $6M_{\oplus}$. The final resonance chain is (2:1, 3:2, 5:4, 4:3).}
    \label{fig:Cold_5Giants}
\end{figure}

\subsection{Outward migration in a cold disc}

{In this section we address the outward migration observed in the cold disc with the four giant planets. A result which contrast with the migration of only Jupiter and Saturn found in \cite{griveaudMigrationPairsGiant2023}.
We aim to explain why the four planets managed to revert their migration direction while the pair did not.}

In $C4$ the arrival of the two ice giants in the common gap pushed the edge of the gap further away from the two gas giants, unbalancing the total torques on the planetary systems. This can be seen in the surface density profile in \cref{fig:Cold_4Giants}, as Neptune arrives in the 5:4 resonance, it brings its own gap with it and thus places the outer edge of the common gap closer to 10 au.
{We also plot in \cref{fig:densProf_outward} the radial density profiles at two moments of the simulation $C4$: when only Jupiter and Saturn are in the disc migrating inwards ($t\simeq 142$ kyr) and when the four planet are migrating outwards ($t\simeq 308$ kyr). These profiles are centered on Jupiter's postion in order to focus on the differences in the gap shape.
As in the \cite{massetReversingTypeII2001} mechanism, the location of the inner and outer Lindblad resonances (resp. ILR and OLR) of each planets play a crucial role in the balance of the total torque. In the case of only Jupiter and Saturn, the gas at the OLR of Saturn apply a strong negative torque on the pair, which in this case is not overcome by the disc at Jupiter's ILR. However, the arrival of the two ice giants significantly decreased the amount of gas present outside of Saturn's orbit, leading to a strong decrease in the negative torque applied on the system.}
{Thus, } in a similar fashion as in the \cite{massetReversingTypeII2001} mechanism, in this case the inner positive Lindblad torque exerted on Jupiter becomes larger than the outer negative Lindblad torque exerted on the other planets. 
This combined with some gas flowing from the outer to the inner disc replenishing the inner edge of the gap, favours {and sustains} an outward migration of the four planets. 

We note that the outward migration observed in $C4$ has stopped once a fifth planet arrived in the system.
The arrival of P9 in $C5$ seems to indicate that this low mass planet has stopped the outward migration of the four giants. 
In fact this is not the case. We show in the top panel of \cref{fig:2panels} the migration of the planets in the simulations $C4$ and $C5$ as well as a case in which instead of removing the gas potential in $C4$, we let the system of four planet migrate longer in the simulation (labelled $C4$ long). This figure shows that the outward migration of the four planets is stopped during the gas disc phase and without the additional fifth planet. 
The effect of P9 is nearly negligible, as should be expected from the mass difference between this planet and the four other giant planets {and the fact that P9 does not perturb the density profile seen in \cref{fig:densProf_outward}.}
The outward migration is eventually stopped when the planets reach an equilibrium radius in the disc, at which a balancing of the inner and the outer Lindblad torques occurs \citep[as also seen for Jupiter and Saturn in][]{morbidelliDynamicsJupiterSaturn2007}. \\

Hence, although we found in \cite{griveaudMigrationPairsGiant2023} that the Jupiter-Saturn pair alone can not reverse its migration and make a `Grand Tack', we see here that in a cold, low-viscosity disc, the ice giants may stabilise the system and prevent Jupiter from migrating inwards and becoming a hot Jupiter. The outward migration may not be fast enough to reproduce \cite{walshLowMassMars2011}'s Grand Tack scenario but it can easily park Jupiter around 5 au at the end of the gas disc phase.

\begin{figure}
    \centering
    \includegraphics[width=\columnwidth]{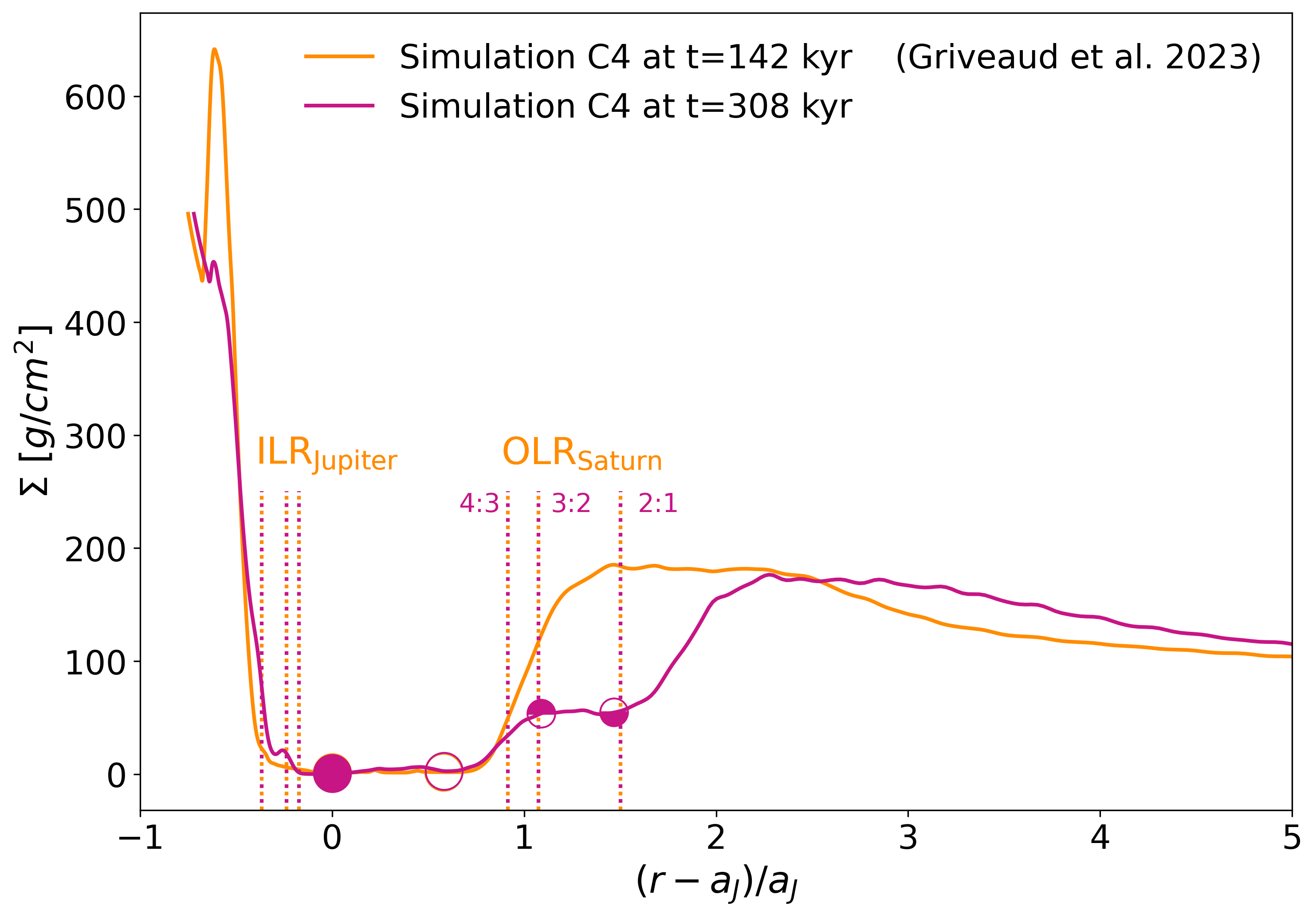}
    \caption{Radial profiles of the surface density in simulation $C4$ rescaled around Jupiter's position. The teal line corresponds to a time when only Jupiter and Saturn are in the disc (essentially the simulation from \cite{griveaudMigrationPairsGiant2023}). At that time, t=142 kyr, the pair of planets is migrating slowly inwards, as seen in \cref{fig:Cold_4Giants}. The magenta line shows the profile when the four giant planets are migrating outwards at t=308 kyr. The planet positons are given by the circle markers. The vertical dotted lines indicate the positions of the inner Lindblad resonances (ILR) of Jupiter and the outer Lindblad resonance (OLR) of Saturn (see text for more details).}
    \label{fig:densProf_outward}
\end{figure}

\begin{figure}
    \centering
    \includegraphics[width=\columnwidth]{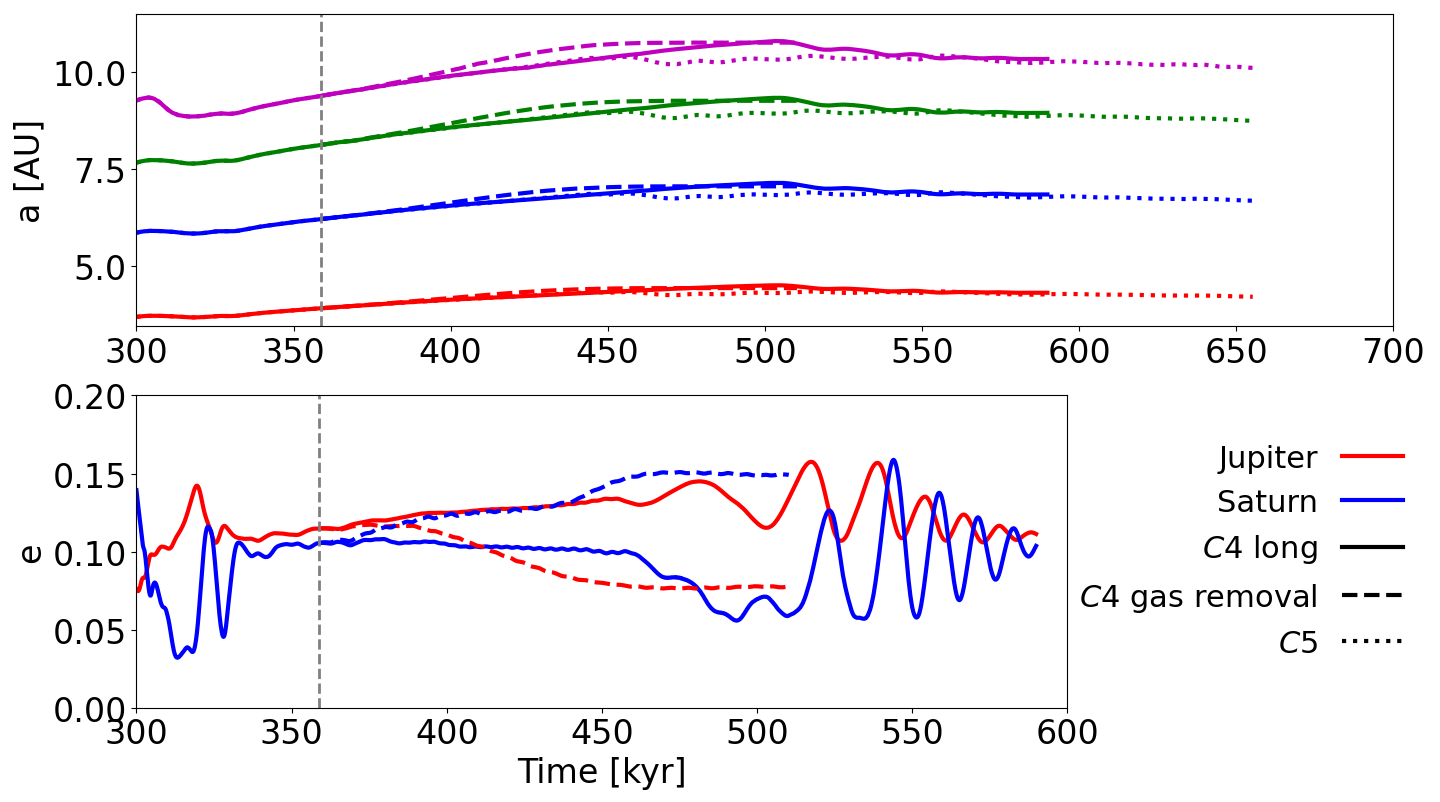}
    \caption{Effect of the gas removal on the planetary orbits in the case of outward migration. Top panel: Semi-major axis evolution of Jupiter, Saturn, Uranus, and Neptune from simulation $C4$ prolonged (plain line), $C4$ with the mock gas evaporation phase (dashed line; corresponding to \cref{fig:Cold_4Giants}), and $C5$ (dotted line; corresponding to \cref{fig:Cold_5Giants} before the mock evaporation phase). 
    Bottom panel: Eccentricity evolution of Jupiter and Saturn in the two $C4$ cases described in the top panel.}
    \label{fig:2panels}
\end{figure}

\subsection{Removing the gas potential}
\label{sec:evap}

The above studied systems are shown to be stable during the gas phase. The presence of gas has a damping effect on the planets' eccentricities which helps the resonant chains to remain stable. However, discs are known to have a physical lifetime of few millions of years, then the gas is dispersed by stellar winds and heating through a photo-evaporation process. In fact, in some studies this gas dispersal can even be the cause of the giant planet instability \citep{liuEarlySolarSystem2022}.
So in order to study the second phase of the giant planets' dynamics in the Solar System, we need to account for the removal of the gas potential on the planets. 
Taking the planets orbital elements directly from the hydro-simulations and setting them in an N-body simulation will create a shock of potential, which could destabilise the system abruptly. Whereas, when the gas is dissipated from the disc, this takes place in a timescale slow enough to let the planets readjust adiabatically their dynamical configuration. In some cases, the system may still go unstable during this phase. Furthermore, we have shown in Appendix B of \cite{griveaudMigrationPairsGiant2023} that the pair Jupiter and Saturn will exchange eccentricity during this gas removal period. It is therefore important to simulate this phase, in order to have the more appropriate orbital elements of the planets before starting our N-body simulation study.

{In the FARGOCA simulations,} we mimicked the effect of gas dispersion by removing smoothly the forces exerted by the disc on the planetary system. 
Precisely, if $\vec a_D$ is the acceleration exerted by the gas onto a planet (including its indirect term, that is the acceleration of the star due to the gas disc), we multiplied it by a function $f(t)$ which smoothly goes from 1 to 0 in the interval $[t_0,t_f]$, with $t_f = t_0+\Delta t$, following
\begin{equation}
    f(t) = \cos^2 \left( \frac {\pi}{2}\frac{(t-t_0)}{\Delta t }\right) \,,
\end{equation}
where $t_0$ is the time at which we start the removal of the disc forces and $\Delta t = 10\,000$ {orbital periods at $r_0$}, corresponding to $\sim 119\,000$ years. \\

This mock evaporation phase is shown in \crefrange{fig:Nom}{fig:Cold_5Giants}, delimited by the vertical dashed line marking $t_0$ and dotted line marking $t_f$. As mentioned above, during this phase, the planets can exchange their eccentricities, this is clearly shown in the bottom panel of \cref{fig:2panels}. In this figure we compare the eccentricities of the planets of simulation $C4$ in the case where they evolve in the gas disc (plain line) and in the case in which we remove the gas potential (dotted line). As soon as this mock evaporation phase is started the eccentricity of Saturn increases while that of Jupiter decreases. This is also seen in simulations $N_{\rm bis}5$ and $C5$. 
The final configurations can now be given to an N-body code as an initial condition to study the next phase of the dynamical evolution of the Solar System.

\begin{table}\centering
\begin{threeparttable} 
    \caption{Exploration of hydrodynamical simulations to form resonant chains with four or five planets.}
    \label{tab:Hydro}
    
    \vspace{0.2cm}
    \begin{tabular}{c c l c}
    $h$      & Name &  Planet order  &  Resonant chain \\
    \hline 
    \multicolumn{4}{c}{4 Planets} \\
    \hline 
    & \\
    0.05       & $N4$            & J-S-U-N   &  (2:1, 3:2, 4:3)      \vspace{2pt}   \\
    0.05       & $N_{\rm bis}4$  & J-S-N-U   &  (2:1, 2:1, 4:3)\tnote{*}     \vspace{2pt}  \\
    0.035          & $C4$            & J-S-U-N   &  (2:1, 3:2, 5:4)      \vspace{5pt}  \\ 
    \hline
    \multicolumn{4}{c}{5 Planets} \\
    \hline
    & \\
    0.05       &                 &  J-S-U-N-U        &  unstable                 \vspace{2pt}  \\
    0.05       &                 &  J-S-U-N-U\tnote{$\dagger$} &  unstable                 \vspace{2pt}  \\
    0.05       &                 &  J-S-U-N-P9       &  unstable                 \vspace{7pt}  \\
    
    0.05       &                 &  J-S-N-U-U        &   unstable                \vspace{2pt}  \\
    0.05       & $N_{\rm bis}5$  &  J-S-N-U-U\tnote{$\dagger$}  &  (2:1, 2:1, 4:3, 5:4)      \vspace{7pt}  \\
    0.035          &                 & J-S-U-N-U         &  unstable                 \vspace{2pt} \\
    0.035          &                 & J-S-U-N-U\tnote{$\dagger$}   &  unstable                 \vspace{2pt} \\
    0.035          & $C5$            & J-S-U-N-P9        &  (2:1, 3:2, 5:4, 4:3)      \vspace{7pt}  \\ 
    0.05       &                 &  J-S-P9-U-N       &  unstable                 \vspace{2pt}  \\
    0.035          &                 & J-S-P9-N-U    &  unstable                 \\ \hline
    \end{tabular}
    \tablefoot{
        This table shows the cases presented in \cref{sec:nominal,sec:cold}, and all cases that did not manage to form a stable resonant chain, to emphasize the difficulty of forming long resonant chains in low-viscosity discs. \\
        Column 1 - Disc aspect ratio $h$. Column 2 - Simulation name if it resulted in a stable configuration. Column 3 - Radial order of the planets in the disc. Column 4 - Final resonant chain at the end of the simulation, if stable.
    }
    \begin{tablenotes}
        \item[$\dagger$] \small{Different introduction time with respect to the above line.}
        \item[*] \small{Stable after having ejected an ice giant.}
    \end{tablenotes}
    \end{threeparttable}
\end{table}

\subsection{Conclusion of the hydrodynamical study}\label{sec:hydro_conclusion}

\Cref{tab:Hydro} summarises the wide exploration of hydrodynamical simulations in attempt to form resonant chains of four or five planets in low-viscosity discs. We obtained three configurations with four giant planets which remained stable throughout the gas phase as well as through the mimicking of the gas dispersal period. In one of them, the system was stable only after having ejected an ice giant in the process. 
We have tested ten configurations with five planets in both nominal and cold disc, varying the initial introduction time, the mass, or even the position of the fifth planet. Only two remained stable during the gas phase and gas dispersal process. 
These simulations show the sensitivity of the system to the time of introduction of the outer planets. It seems that in low-viscosity discs, the timing or the order in which the planets enter in resonance with one-another plays a more important role in the final structure of the system. In such discs, the gaps carved by giant planets are much deeper and wider than in classically viscous discs. Consequently, there is less dissipation around the planets and therefore the resonance chains are weaker.

Analysing the resonant chains obtained in this work, we find firstly that Jupiter and Saturn remain locked in the 2:1 MMR despite the addition of multiple planets in the system. This confirms the robustness of the result found in \cite{griveaudMigrationPairsGiant2023}. This leads to having chains with a wider separation than in classically viscous discs by construction, since we recall at higher viscosity Jupiter and Saturn are most often found in the 3:2 MMR. For the resonances between the other planets we find the usual first order resonances often found in such systems, that is the 3:2, 4:3 or 5:4 \citep[as seen in][]{morbidelliDynamicsGiantPlanets2007,nesvornySTATISTICALSTUDYEARLY2012}. In the specific case where an ice giant is ejected, this allows Neptune to be captured in the 2:1 MMR with Saturn, creating an even wider chain than previously seen. 
Only one of our resonant chains has been studied previously in the context of a Nice Model-like study by \cite{nesvornySTATISTICALSTUDYEARLY2012} and \cite{thommesMeanMotionResonances2008} which was found in our simulation $N4$. 

{Except for \cite{morbidelliDynamicsGiantPlanets2007},} the only comparison we have with the building of multiple giant planets configuration in high-viscosity discs comes from studies in which the migration is ran in N-body simulations with a prescription for the gas' potential \citep[as seen in e.g.][]{nesvornySTATISTICALSTUDYEARLY2012, pierensOutwardMigrationJupiter2014, clementBornEccentricConstraints2021}. 
In order to mimic migration in a disc, a damping force is applied to the semi-major axis and eccentricity of the planets based on estimates from analytical migration rates {or calibrated from high-viscosity hydrodynamical simulations \citep[e.g.][]{cresswellThreedimensionalSimulationsMultiple2008}.} The damping of the eccentricity is often set to be a factor $K$ times stronger than that of the semi-major axis, where $K\sim 100$ \citep{leeDynamicsOriginOrbital2002,batyginEarlyDynamicalEvolution2010}. 

However, recent studies \citep{pichierriRecipeEccentricityInclination2024} suggest that because of the larger gaps carved in low-viscosity discs, K is most likely lower. Reducing $K$ would also substantially reduce the number of stable resonant chains obtained in N-body simulations \citep{bitschEccentricityDistributionGiant2020a,batyginDissipativeCapturePlanets2023} {and could help to reproduce eccentricity distributions of Cold Jupiters \citep{bitschEccentricityDistributionGiant2020a,matsumuraNbodySimulationsPlanet2021}}. {Most importantly, in general a common K coefficient} is applied to all the planets in the system, but {K should depend on the individual gap shapes, which in turn depends on the planet masses and local disc parameters. Hence, the four giant planets} will evidently not feel the same damping from the gas disc. 
{This complexity thus} highlights the importance of hydro-simulations versus N-body with prescription. 
{Lastly, simulating the transition between hydro-simulations and N-body is important, as} we have also seen in \cref{sec:evap} that planets can exchange eccentricities during this transition phase.

Finally, {an important result of this section is the reversal of the migration direction of the four giant planets} in a cold protoplanetary disc. It seems that a similar mechanism to the two-planets Masset \& Snellgrove mechanism enters into play when the system has two massive gas giants followed by two less massive ice giants {in a low-viscosity disc}. 
This mechanism is a great solution to stop Jupiter's inward migration and thus solve the problem unveiled by \cite{griveaudMigrationPairsGiant2023} for the formation of the Solar System.


\section{Giant planet instability: N-body study}\label{sec:Nbody}

Once we obtained five stable resonant chains between 4 or 5 planets, we are interested in studying whether these resonant configurations can recreate the Solar System after some interactions with a planetesimal disc, similar to the previous Nice model studies \citep[e.g.][]{tsiganisOriginOrbitalArchitecture2005,nesvornySTATISTICALSTUDYEARLY2012}. 

\subsection{Methods}\label{sec:Nbody_methods}

To study the N-body interactions between the planets and the planetesimal discs, we used the N-body library \texttt{rebound} \citep{reinREBOUNDOpensourceMultipurpose2012}. 

We initialised the positions of the planets using the orbital parameters outputted by the code FARGOCA after the mock disc evaporation phase described in section \ref{sec:evap}. For the purpose of comparison with the Solar System we rescaled the semi-major axes of the planets such that Jupiter is located at about $5.2$ au {at this stage; technically, this is just a change of length unit}. We emphasise that the result of the migration in the gas phase is independent of the scaling of the simulation (see Sect. 2.2 of \cite{griveaudMigrationPairsGiant2023}). 
We initialised the position of each planet with the semi-major axis, eccentricity, mean longitude and longitude of pericentre given by FARGOCA. However, because the hydrodynamical simulations presented in \cref{sec:hydro} are two-dimensional, the planets come out to be co-planar. In order for them to evolve in three-dimension in this part of the study, we initialised the planets with an inclination 
$\sin i = |X|$ where $X$ is randomly drawn from a Gaussian distribution with a mean of zero and standard deviation of $10^{-3}$, and a longitude of ascending node randomly determined from a uniform distribution between $0$ and $2\pi$.

We added a ring of planetesimals outside the last planet's orbit. Planetesimals are implemented in the simulation as `small particles' (i.e. while they interact with the planets) the gravitational interactions between the particles are neglected. Naturally, the planets interact with each other and with the disc. This is a well accepted assumption made in the previous Nice model studies, since the main driver of the giant planet instability comes from the interactions between the planets themselves and from their interactions with the planetesimal disc.  

In our work, the planetesimals disc is composed of $2000$ super-particles and is located between $r_{\rm min} = r_{p,\rm ext} + 0.5$ au and $r_{\rm max} = 30$ au, where here $r_{p,\rm ext}$ is the position of the last planet in the resonant chain. {The value of $r_{\rm min}$ only affects the timing of the instability but not the outcome as also found by \cite{nesvornySTATISTICALSTUDYEARLY2012}. In order to have a quick instability and save computational time, we chose a short distance of $r_{\rm min}$.} The particles were radially distributed linearly within this interval {resulting in a surface density of the planetesimal disc in $r^{-1}$}. Each particle was given an initial random eccentricity and inclination of the order of $10^{-4}$. All particles have the same mass given by the total mass of the disc, $M_d$, divided by the number of particles. In this work, we explored two different disc masses: $30 M_\oplus$ and $50 M_\oplus$.

The system was then integrated using the hybrid symplectic integrator \texttt{Mercurius} \citep{reinHybridSymplecticIntegrators2019}, with an integrator time step of about a fiftieth of the most inner planet's orbital period. The system was generally integrated up to about a billion year, except when a system was left with less than four planets, then the simulation was stopped. During the integration, particles {or planets} {reaching} beyond 1000 au were removed from the simulation.

Finally, because the outcome of a global instability is very sensitive to initial conditions, we ran a statistical study of hundreds of simulations for each resonant chain obtained from the hydro-simulations by varying the initial positions of the planetesimals. We show the result of this study in the following section.

\subsection{Assessment of success}\label{sec:criterions}

In order to quantify how well a simulation can reproduce the Solar System, we used the criteria described in \cite{nesvornySTATISTICALSTUDYEARLY2012}. Although these criteria have been adapted in more recent studies \citep{deiennoConstrainingGiantPlanets2017,clementBornEccentricConstraints2021} to better fit the constraints of the Solar System, for the scope of this work we used the original versions of the criteria (with an even more simplified criterion C, hereafter called C$^\ast$). We recall here their definition below.

\paragraph{{Criterion A}}
The first obvious feature to reproduce is to have four giant planets left at the end of the simulation{, in an order comparable to the current one, that is Jupiter inner to Saturn, inner to two ice giants. We consider the two ice giants to be interchangeable and therefore name them according to their final radial order.}

\paragraph{{Criterion B}}
The next important constraint is to match as much as possible their current orbits. For this, each planet must be within $20\%$ of the current semi-major axis of the planet and for their eccentricities to be not higher than $0.11$ and their inclinations less than $2^{\circ}$. The inclination is defined with respect to the invariant plane of the system, that is the plane perpendicular to the total angular momentum of the system.

\paragraph{{Criterion C}}
The eccentricities of the giant planets, especially that of Jupiter, are an important feature in today's dynamics of the Solar System.
But Jupiter's dynamics is not just about its eccentricity $e_5$,\footnote{{We follow the traditional notation from the Solar System secular studies where the giant planets are numbered from 5 to 8 even if we don't simulate the terrestrial planets dynamics.}} but also about which secular frequencies are dominant. To reproduce today's system, the amplitude of the eccentric mode $e_{55}$, should be at least half of its current value so $e_{55}>0.022$ \citep{nesvornySTATISTICALSTUDYEARLY2012}.
In addition, a large $e_{55}$ indicates a close encounter with an ice giant, which is also necessary for capture of the irregular satellites and trojan asteroids.

To start with, $e_5$ being essentially the sum of the Fourier modes associated to the four giant planets, $\sqrt{\langle e_5(t)^2 \rangle} = \sqrt{\sum_{k=5}^8 |e_{5k}|^2} \geq e_{55}$. Therefore, to avoid calculating $e_{55}$ for all our simulations, we adopt a pre-criterion, named hereafter criterion C$^\ast$, simply requiring:
$$\sqrt{\langle e^2 \rangle} \geq e_{55} \geq 0.022\,,$$ where $\langle \cdot \rangle$ marks the temporal average. 

If this criterion is respected, we then calculate the amplitude $e_{55}$ by integrating the system without any planetesimals and with a smaller time-step and conduct a Fourier analysis the result. 

\paragraph{{Criterion D}}
Finally, the last criterion requires the period ratio between Jupiter and Saturn, $P_{\rm S}/P_{\rm J}$, to evolve from $<2.1$ to $>2.3$ in less than $1\, \rm Myr$ and the final $P_{\rm S}/P_{\rm J} < 2.8$. This criterion insures the survival of both the terrestrial planets \citep{brasserConstructingSecularArchitecture2009, agnorMigrationJupiterSaturn2012} and the asteroid belt \citep{morbidelliEvidenceAsteroidBelt2010,mintonSecularResonanceSweeping2011}. This criterion was set in the context of a late instability. In this context the terrestrial planets would already be formed when the instability triggers, and thus it was important to have a criterion assessing their survival. However, in our simulations we observe than in the cases where an instability is triggered, it happens within the first 10 Myr. Unless the terrestrial planets would be formed at this time, we could thus relax this criterion. \\
Actually, \cite{clementBornEccentricConstraints2021} consider D as simply having $P_{\rm S}/P_{\rm J} < 2.5$. However, we also need Jupiter and Saturn to have migrated a minimum, so we redefine a new criterion named D$^\ast$ requiring $2.3 < P_{\rm S}/P_{\rm J} < 2.5$, and we will show both versions in our results. 

It should be noted  that criteria C and D   both include criteria A and B, but are assessed independently of one another.


\begin{table*}\centering
    \caption{Results of N-body  simulations integrating the four or five giant planets initially in a resonant chain built in \cref{sec:hydro}.}
    \label{tab:NBody}
    
    \vspace{0.2cm}
    \begin{tabular}{l l l c c c c c c}
      Resonant chain  & Name & $N_{\rm sim}$ & $M_D$           &  A  &  B &  C (C$^\ast$) &  D; D$^\ast$    &  C$\cap$D; C$\cap$D$^\ast$    \\
                     & &  & ($M_\oplus$)    & (\%)  & (\%) & (\%)  & (\%)     &  (\%)  \\

    \hline
    \multicolumn{9}{c}{Cold protoplanetary disc ($h=0.035$)} \\ 
    \hline
    \multirow{2}{*}{(2:1, 3:2, 5:4)}   & \multirow{2}{*}{$C4$} & 200   & 30 &   7    &  0.5   &  0.5   \, (0.5)       &  0 \,; 0   &     \\ 

                        &      & 200   & 50 &  25.5  &  18    &  1     \, (4)       &  3 \,; 10   &   0 \,;\, 1  \vspace{4pt}\\  

    \multirow{2}{*}{(2:1, 3:2, 5:4, 4:3)} & \multirow{2}{*}{$C5$} & 200   & 30 &  9     &  4   &  1     \, (1)          &  0 \,; 0   &     \\
                         &      & 200   & 50 &  21    &  13.5  & 1.5    \, (3)    &  3 \,; 4   &   0.5 \,;\, 1.5  \vspace{7pt}\\

                         \hline
    \multicolumn{9}{c}{Nominal protoplanetary disc ($h=0.05$)} \\ 
    \hline 
    \multirow{2}{*}{(2:1, 3:2, 4:3)} & \multirow{2}{*}{$N4$}            & 50   & 30  & 98 & 1 & 0 (0) & 0\,;\,0  &      \\
                        &                 & 100   & 50  & 69 & 29 &  1 (2) &  2\,;\,10  &   1\,;\, 0  \vspace{4pt}\\

    \multirow{2}{*}{(2:1, 2:1, 4:3)} & \multirow{2}{*}{$N_{\rm bis}4$}  & 50   & 30  & 100 & 100 & 0 (0)  & 0 \,; 0  &     \\
                        &                 & 50   & 50  & 100 & 96  & 0 (0)  & 0 \,; 70  &    \vspace{4pt} \\

    \multirow{2}{*}{(2:1, 2:1, 4:3, 5:4)} & \multirow{2}{*}{$N_{\rm bis}5$} & 100 & 30 & 19 & 10  & 0 (2) \,   & 3 \,; 0  &     \\
                        &                     & 100 & 50 & 4  & 3   & 0 (0)  &  1 \,; 1  &    \vspace{4pt}\\
    \end{tabular}
    \tablefoot{
        Column 1 - Initial resonant chain. Column 2 - Name of the hydrodynamical simulation. Column 3 - Number of N-body simulations ran for the statistical study. Column 4 - Mass of the planetesimal disc placed outside of the last planets' orbit. Column 5-9 - Success rate of the criteria detailed in \cref{sec:criterions}.
        }
\end{table*}

\subsection{Results}\label{sec:Nbody_results}

In this section, we show the results of the N-body simulations ran for each resonant chain. Although the protoplanetary disc's property does not matter any more, in this section of the work, we still categorise the results by the aspect ratio of the disc in which the resonant chain was formed. The planetary systems coming from the same protoplanetary disc seem to show similar trends in our results.

\subsubsection{Chains forming in a cold protoplanetary disc}

In this section we let the planetary systems that formed in a cold protoplanetary disc, simulations $C4$ and $C5$ which resulted in the resonant chains (2:1, 3:2, 5:4) and (2:1, 3:2, 5:4, 4:3) respectively, evolve in the presence of a planetesimal disc. 

Though first, we start by assessing the stability of our resonant chains without the planetesimal disc. 
While varying the initial inclinations and longitude of ascending node of the planets (which we recall are randomly determined), we find that the $C4$ chain goes unstable up to 4.5 Myr (in 10 simulations). $C5$ becomes unstable even faster, losing at least 2 planets in about a million year or less. 
Now adding the disc of planetesimals allows some systems to remain stable for longer, as the planetesimals act as a damping force on the planets orbital changes \citep{stewartEvolutionPlanetesimalVelocities1988}. \Cref{tab:NBody} shows the results obtained for both resonant chains and two planetesimal disc masses. Since these chains were highly unstable, in order to increase our statistics we ran 200 simulations for each configuration. 

\paragraph{Chain (2:1, 3:2, 5:4)}
The case with a disc of $M_D = 30M_\oplus$ yields to very poor passing of the criteria. The system is highly unstable and loses a planet in $93\%$ of the cases leading to a very poor result of criterion A. The disc of planetesimals is not massive enough to damp the planets excited orbits leading to at least one of them being ejected out of the system. This case is not a likely outcome for the Solar System. However, if the planetesimal disc mass is of $50M_\oplus$, this leads to much more simulations passing through all four criteria. The system still goes unstable very early ($\lesssim 10^6$ yr) as in the case of $M_D = 30M_\oplus$, but the planetesimals manage to damp the planets eccentricities and inclination in order to maintain them in long-term stable orbits in $25.5\%$ of the cases. The results of these simulations are shown in \cref{fig:stat_4C_50EM}, where the top (resp. bottom) panel show the eccentricities (resp. inclinations) of the planets as a function of their semi-major axes. \\
Among those $25.5\%$, in $18\%$ of the simulations the planets final positions pass B, which is represented by the coloured rectangles in the figure. Not only do they pass this criterion, but the mean value of the distribution of each planet is very close to the positions of the planets today, both for the eccentricity and inclinations. \\
In a case with four planets, the eccentricity of Jupiter is arduously met, as the gas giant cannot benefit from the ejection of a planet to boost its eccentricity. Most of those $18\%$ yield a too low eccentricity of Jupiter, in fact only $4\%$ have an eccentricity significant enough to qualify for C (so that pass criterion C$^\ast$). Among those, only 1\% actually validate that criterion. This shows that simply checking whether Jupiter's eccentricity is high, is not enough to verify whether this simulation can reproduce the finer aspects of the Solar System's dynamics. In some cases the eccentricity of Jupiter is boosted by the high eccentricity of Saturn, in such way that the dominating secular mode is $e_{56}$ and not $e_{55}$. These cases thus pass our pre-criterion C$^\ast$ but not C. {Importantly, we checked that in the ones that do validate C, the eccentricity of Jupiter is triggered by a close encounter with an ice giant. In these case encountered planets is maintained in the system by the massive planetesimal disc.}\\
With this resonant chain, the instability is triggered very early (within $<10$Myr) so the passage of D. is not a strong constraint. Nevertheless, about 3\% pass the conservative criterion D. If we consider the looser criterion D$^\ast$ the success rate reaches 10\%. For this chain we obtain 1\% of successful run passing simultaneously C and D$^\ast$.

\begin{figure}
    \centering
    \includegraphics[width=\columnwidth]{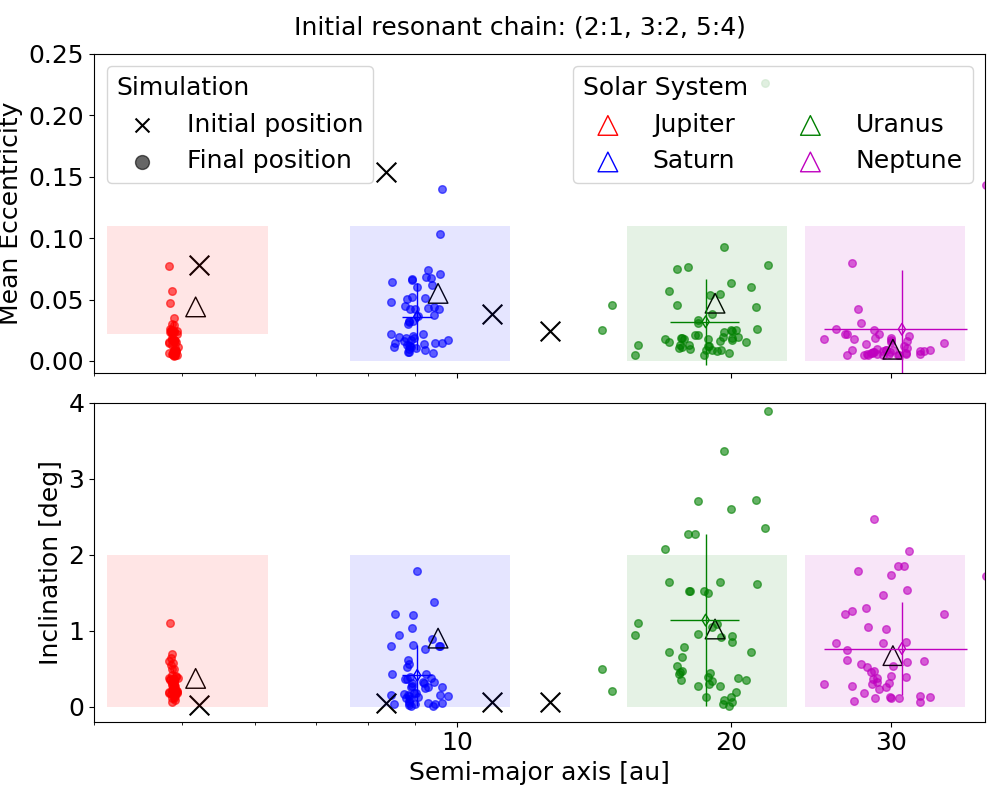}
    \caption{Results of the N-body integrations of 200 simulations starting from the chain (2:1, 3:2, 5:4) (simulation $C4$ in \cref{sec:hydro}) with a disc of $50 M_\oplus$. 
    The figure shows the mean eccentricities (\textit{top panel}) and inclinations (\textit{bottom panel}) of the planets with respect to their semi-major axes. 
    The initial positions of the planets are given by the black crosses and their positions after integration are shown by the coloured dots. The coloured bars give the mean and standard deviation of the distribution of orbital parameters for each planet. Only the simulations passing criterion A are shown in the figure (here $25.5\%$). The coloured boxes represent criteria B and C$^\ast$ (see text for more details).
    For reference, the current positions of the Solar System planets are given by the triangles.
    }
    \label{fig:stat_4C_50EM}
\end{figure}

\paragraph{Chain (2:1, 3:2, 5:4, 4:3)}
With an additional fifth planet of $6M_\oplus$, namely P9, we find similar trend between the different cases of disc mass. With a disc of $30M_\oplus$, the planetesimals again are not able to damp the planets high eccentricities and inclinations leading to the loss of at least two planets in about 66\% of the times. In many cases ($\sim 23\%$), P9 remains in the system while one of the ice giants is ejected such that the simulation does not validate criterion A. 
However, when increasing the mass of the disc to $50M_\oplus$, we find a better success rate of the criteria. We show those successful runs in \cref{fig:stat_5C_50EM}. With such resonant chain we manage to have at least 1.5\% of the simulations validating up to criterion C and 4\% passing D. In this configuration we obtain 1.5\% of the simulations which combines criteria C and D$^\ast$ simultaneously. \\

Resonant chains built in a cold protoplanetary disc are good candidate to form the current Solar System. They are easily destabilised and create early instabilities. How likely/often do they reproduce the Solar System in our study depends mostly on the selected criteria. If the terrestrial planets were already formed at the end of the protoplanetary disc phase {or if the instability is actually late\footnote{We recall our initial conditions trigger an early instability for numerical reasons.} \citep{avdellidouDatingSolarSystem2024,izidoroLinkAthorMeteorites2024}}, a simulation must validate the strict criterion D. In this case, the resonant chain of five planet (2:1, 3:2, 5:4, 4:3) with a planetesimal disc of 50 Earth masses is capable of yielding such result, with a small success rate admittedly, but a success nonetheless. However, {there is some evidence} that {the final assembly of} the terrestrial planets {could take place} after at least 40 Myr \citep[e.g.][]{kleineHfWChronologyAccretion2009,jacobsonHighlySiderophileElements2014}, and because these resonant chains go unstable within the first 10 Myr after the gas disc's dispersal, we can consider criterion D$^\ast$ enough. In this case both resonant chains with four or five planets, formed in a cold protoplanetary disc, can reproduce the Solar System with $\sim 1\%$ success rate.

\begin{figure}
    \centering
    \includegraphics[width=\columnwidth]{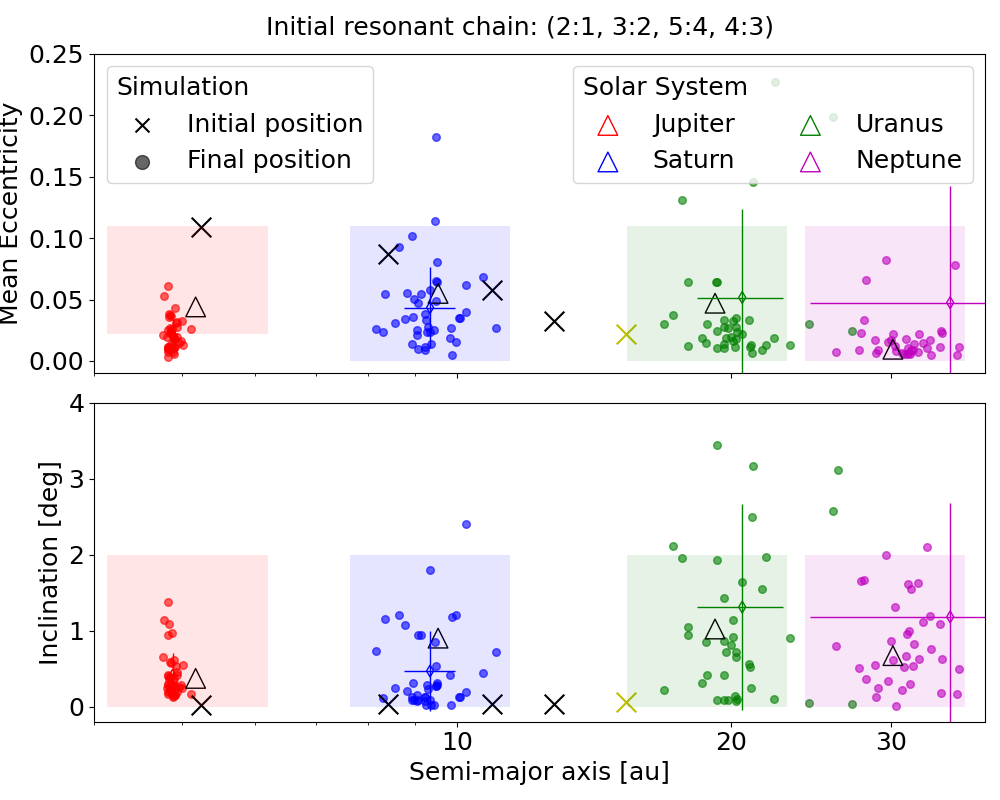}
    \caption{Similar to \cref{fig:stat_4C_50EM}, but for the initial chain (2:1, 3:2, 5:4, 4:3) (corresponding to $C5$ in \cref{sec:hydro}), with a disc mass of $50 M_\oplus$. The yellow cross marks the initial position of the fifth planet, P9, which must get ejected in order for the simulation to pass criterion A.} 
    \label{fig:stat_5C_50EM}
\end{figure}

\subsubsection{Chains forming in a nominal protoplanetary disc}

Resonant chains forming in the nominal protoplanetary disc (so $h=0.05$) with four planets are stable for over 100 Myr without any planetesimals. The one with five planets tends to become unstable and ejects two or more planets in a time span of 1-10 Myr.

\paragraph{Chain (2:1, 2:1, 4:3)}
This resonant chain leads to a large success rate of A and B but totally fails C. and D. In this case, the planets follow a smooth planetesimals-induced migration to their current positions but the system never goes unstable, meaning there are no close encounters between planets. Their semi-major axes simply diverge slowly during the evolution of the system. We show an example of a typical simulation in \cref{fig:Nbis4_orbel}.
\Cref{fig:stat_Nbis4} shows the statistical results of the all simulations ran for both disc masses. The only notable difference between both is that a more massive disc manages to spread the planets further apart. 
This smooth migration gives good estimates of the semi-major axes of the planets, but it does not provide any boost in the eccentricities or inclinations. 
Although our hydro-simulations give a high eccentricity to Jupiter and Saturn during the disc phase, as soon as the planets start interacting with the disc of planetesimals, their eccentricity is damped to nearly zero. From there some close encounters between planets are required to re-excite them to their current values.  This result was also observed by \cite{nesvornySTATISTICALSTUDYEARLY2012} in the cases where they studied wide resonant chains or chains starting with the 2:1 MMR.

\paragraph{Chain (2:1, 2:1, 4:3, 5:4)} 
Despite being easily destabilised without planetesimals, this resonant chain remains stable more often in the presence of a planetesimal disc. For $M_D = 30M_\oplus$, 67\% of the simulations keep five planets, which follow a smooth migration as seen with the chain (2:1, 2:1, 4:3), while this number goes up to 93\% for $M_D = 50M_\oplus$. Unlike the cases of chains forming in a cold protoplanetary disc, this chain shows more favourable results with a lower mass planetesimal disc. In this case a massive disc favours a smooth migration, while a less massive disc allows the planets to get excited enough for the system to go unstable about a third of time. However, those that do succeed in passing A and B do not have Jupiter's eccentricity excited enough to pass C. Regarding D, 3\% of the simulations pass D, but all have a final $P_{\rm S}/P_{\rm J} > 2.5$ and thus do not pass D$^\ast$. Overall even though this resonant chain undergoes some instabilities, it is not enough to successfully reproduce the Solar System.\\

\paragraph{Chain (2:1, 3:2, 4:3)}
This chain was already studied in \cite{nesvornySTATISTICALSTUDYEARLY2012}. They found that the planets follow a smooth migration which does not excite $e_{55}$ enough nor satisfies D. In our study, even though we have the same resonant chain, the full orbital structure of the planetary system might not be exactly as the one studied in \cite{nesvornySTATISTICALSTUDYEARLY2012} for reason explained in \cref{sec:hydro_conclusion}. In fact, we do not find the same behaviour nor a systematic smooth migration as in the cited study. For a massive planetesimal disc, the system goes unstable and loses a planet in at least one third of the times. In the remaining cases, {29\%} pass B and few percent pass C and D. Overall for $M_D = 50M_\oplus$, we obtain a 1\% success of all criteria. {However, in the simulation that does pass C, we find that Jupiter does not have any close encounter with another planet. Thus even though it passes all the right criteria, it is not a likely reproduction of the Solar System.}
With a less massive disc though, the planets hardly migrate. The system barely gets unstable ($<1\%$) and we do not even observe a smooth migration as for the chain (2:1, 2:1, 4:3). The system ejects planetesimals at a very low rate. This case is not a good choice for reproducing the Solar System.\\

Chains forming in a protoplanetary disc with scale height of $h=0.05$ are much more stable than those forming in a thinner disc. Those starting with (2:1, 2:1, ...) do not yield any success of reproduce the Solar System. These chains are highly stable and the interactions with a disc of planetesimals are not enough to make the system undergo an instability. There is a systematic failure of criteria C and D. In the case were the innermost ice giant is closer to Saturn, so (2:1, 3:2, 4:3), the success depends strongly on the mass of the planetesimal disc. If too light, the disc does not affect the planetary system. If massive enough, the disc can trigger an instability which leads to a 1\% success rate of reproducing the Solar System.

\begin{figure}
    \centering
    \includegraphics[width=\columnwidth]{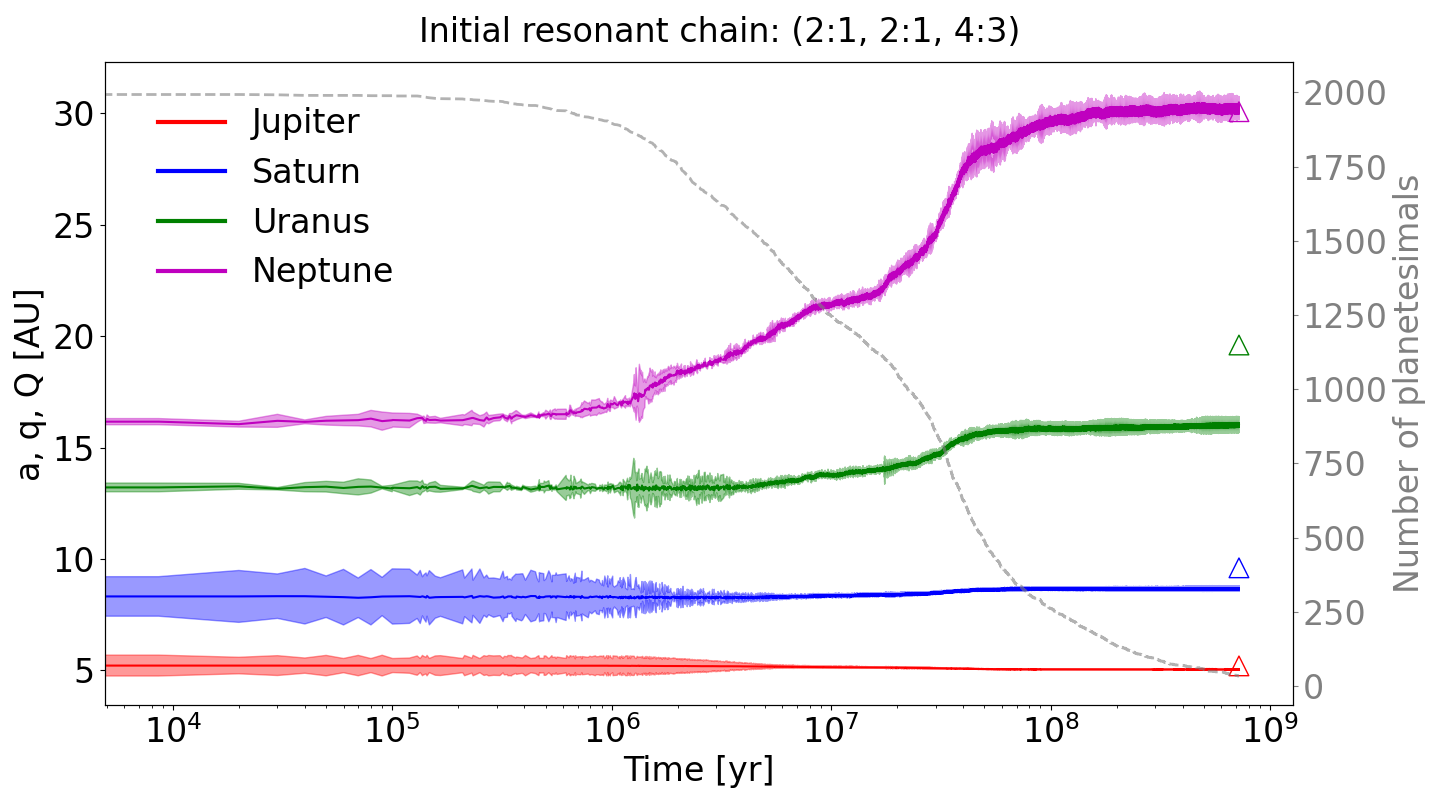}
    \caption{Evolution of the orbital parameters of the planets in one simulation of the chain (2:1, 2:1, 4:3) and $M_D=30M_\oplus$, to illustrate the smooth planetesimal-driven migration of the planets. The dashed grey line in the background shows the number of planetesimals in the simulation.}
    \label{fig:Nbis4_orbel}
\end{figure}

\begin{figure}
    \centering
    \includegraphics[width=\columnwidth]{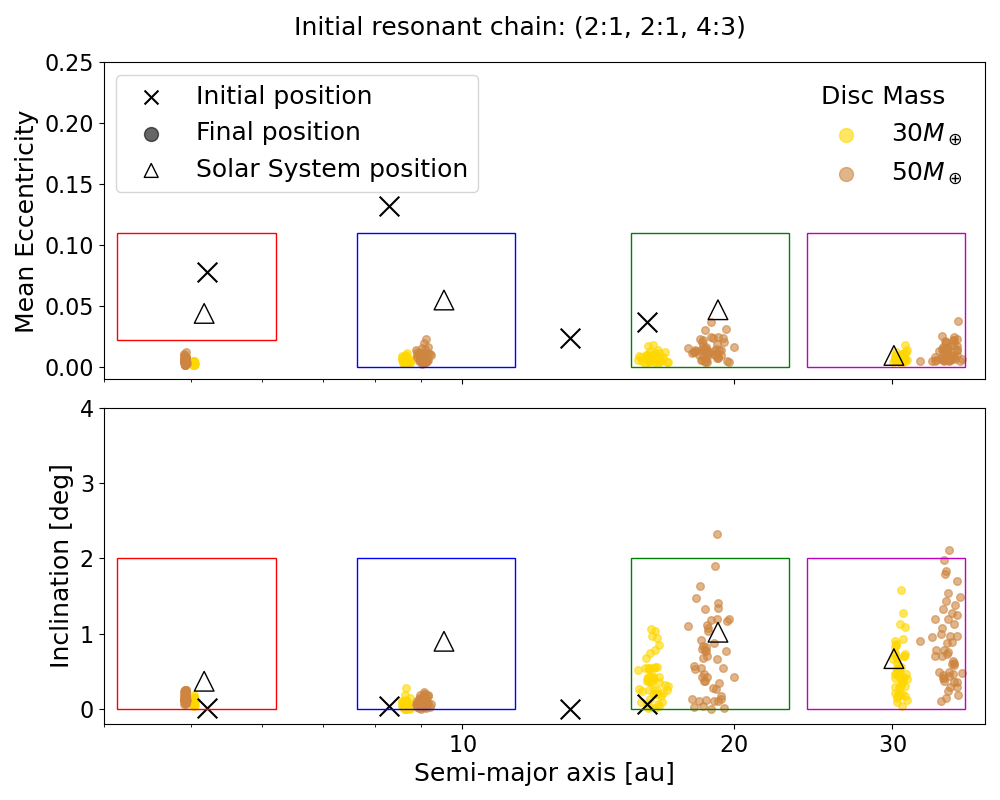}
    \caption{Similar to \cref{fig:stat_4C_50EM}, but for the result of simulations starting from the chain (2:1, 2:1, 4:3), so $N_{\rm bis}4$. In this case the results are shown for both disc masses: in light yellow for $M_D=30M_\oplus$ and in darker brown for $M_D=50M_\oplus$. The eccentricity of Jupiter is systematically too low to yield any success at reproducing the Solar System.}
    \label{fig:stat_Nbis4}
\end{figure}

\section{Discussion and conclusion} \label{sec:conclusion}

\begin{figure*}
    \centering
    \includegraphics[width=\textwidth]{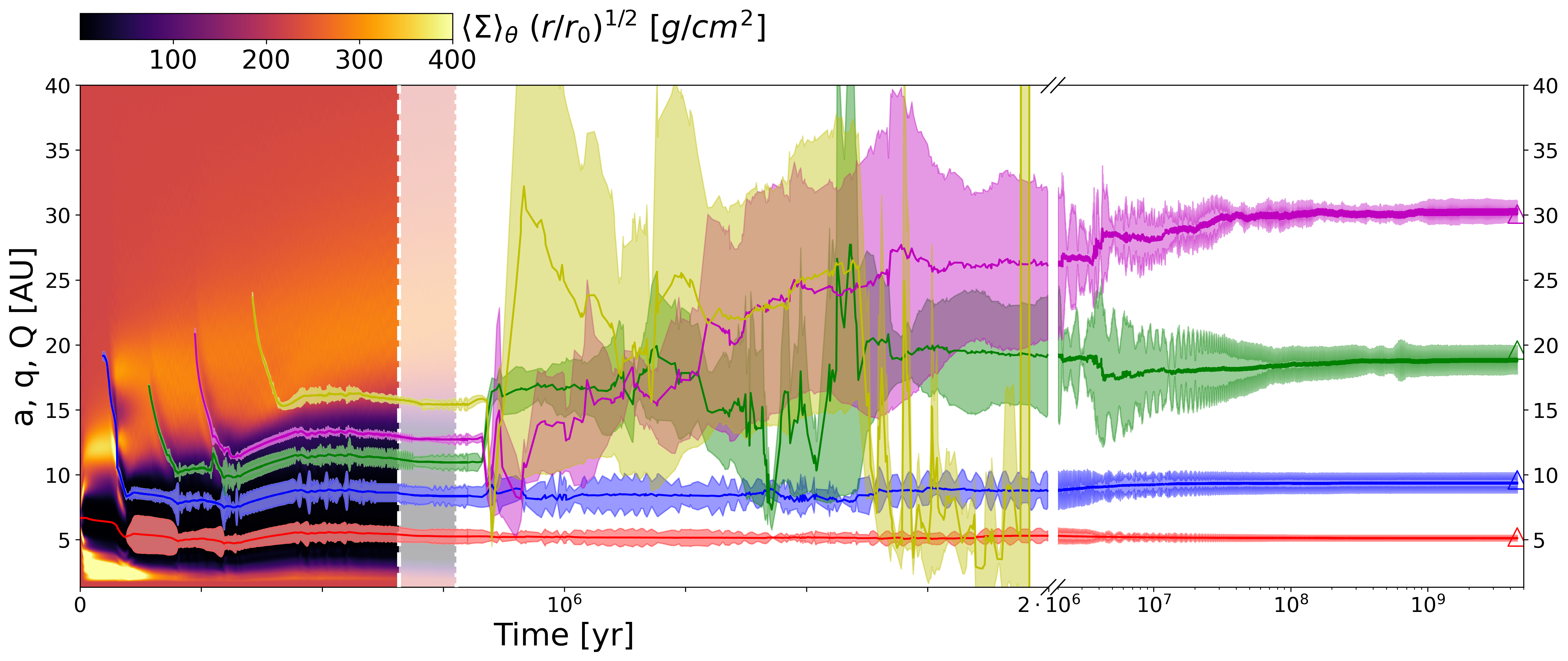}
    \caption{Possible scenario of the dynamical history of the four giant planets of our Solar System if they were to form in a cold low-viscosity protoplanetary disc and then interact with a $50M_\oplus$ planetesimal disc located at 0.5 au away from the last planet. In this scenario the initial planetary system is composed of five planets, where the fifth planet is a $6M_\oplus$ planet that could correspond to P9.  
    The first 800 kyr show a hydrodynamical simulation of the protoplanetary disc phase, in which the planets migrate into a resonant configuration. In the background we show the radial profiles of the gas surface density in the disc. We then simulate a mock gas-dispersal period (between the two white dashed lines). Finally, we show the result of one N-body simulation in which the planets interact with a planetesimal disc. The system goes unstable quickly and ejects the fifth planet before 2 Myr. We then integrate the system for a few billion years (with a change in scaling of the x-axis from linear to logarithmic) and obtain a good analogue of the present Solar System (the triangles mark the actual semi-major axes of the giant planets).}
    \label{fig:final}
\end{figure*}

The aim of this work was to retrace the dynamical history of the giant planets of the Solar System from the protoplanetary disc phase until the Nice Model. We accomplished this by first building resonant chains using fully hydrodynamical simulations, then after a mock disc evaporation phase, we integrated the planets' interactions with a disc of planetesimals placed outside the last planets orbit. Such a fully consistent study had not been done since the combined studies of \cite{morbidelliDynamicsGiantPlanets2007} and \cite{levisonLATEORBITALINSTABILITIES2011} in the so-called Nice Model 2. However, these were done in the context of a high-viscosity protoplanetary disc on the one hand, and with the requirement of a late instability for the Nice Model on the other hand. Since then the protoplanetary disc paradigm has shifted towards low-viscosity discs and the constraint on the timing of the instability has been lifted. 
It was therefore necessary to run a fully consistent study of the possible formation of the Solar System in this new context. 

To start with, we ran two-dimensional hydrodynamical simulations with a viscosity parameter of $\alpha=10^{-4}$, studying the convergent migration of giant planets. We had already shown in \cite{griveaudMigrationPairsGiant2023} that the Jupiter-Saturn pair would never get past the 2:1 MMR in such a low-viscosity disc. We added the two ice giants in the system and found three different resonant chains for two different disc scale heights. Despite the adversity, we also obtained two resonant chains with five planets. Finally, in order to study the Nice Model phase, we mimicked the evaporation phase of the protoplanetary disc allowing the planetary system to readjust adiabatically its configuration. 

The first half of this work shows the importance of hydrodynamical studies in order to build resonant chains. Except for the above-cited study, all other Nice Model studies have used resonant chains built using N-body integrators with either a prescription of migration or a simple damping force on the eccentricities and semi-major axes of the planets. We show in this work that building resonant chains in hydrodynamical simulations can be  much more complex. The stability of the chain depends highly on the distribution of the gas around the planets. In such a low-viscosity disc, giant planets create wide and deep gaps, which reduce significantly the dissipation forces of the gas on the resonances. Therefore, systems of multiple giant planets in low-viscosity discs are easily unstable, especially for more than three planets. 

The five resonant chains we obtained are wider than those previously studied in the Nice Model studies, because the Jupiter-Saturn pair is always in the 2:1 MMR. In two cases, the second pair is also in the 2:1 MMR and in the 3:2 in the other chains. 
While previous studies considered Jupiter and Saturn in the 2:1 MMR mainly for academic purposes, some also considered the second pair in the 2:1 MMR and concluded that these systems were highly stable \citep{nesvornySTATISTICALSTUDYEARLY2012}. Our results are in agreement with this statement as our simulations $N_{bis}4$ and $N_{bis}5$ show barely any instabilities. 
If the second pair of planets is in the 3:2 MMR, the literature also shows that these chains generally do not lead to any instabilities \citep{deiennoConstrainingGiantPlanets2017, clementBornEccentricConstraints2021, nesvornySTATISTICALSTUDYEARLY2012}. Our results show that if the chain is built in a cold PPD, the system is highly unstable when interacting with a planetesimal disc. A $30M_\oplus$ disc is not enough to maintain all the planets in the system, while a $50M_\oplus$ disc gives successful results, also in the case of four or five planets. However, if the chain forms in the nominal PPD, the system will be more stable, although some instabilities can be obtained in the case of a $50M_\oplus$ planetesimal disc. 
We think that resonant chains forming in a cold protoplanetary disc are weakly stable due to the lack of gas damping effects in the planets' vicinity, and therefore can  explain our results and possibly the difference with the literature. 
In the low-viscosity context, it is realistic to consider protoplanetary discs with low aspect ratios similar to that of our cold disc. We also note that in such discs, we found that the four giant planets migrate outwards during the protoplanetary disc phase, thus preventing them from becoming hot Jupiters.

Among all the systems we have studied, three configurations satisfy the four criteria required to fit the constraints of the Solar System with about 1\% of success each, {which is slightly lower than the best cases of \cite{nesvornySTATISTICALSTUDYEARLY2012}, although of the same order of magnitude.} {In \cref{fig:final} we show an example of such an evolution, from the migration in the protoplanetary disc to the present.}
The Solar System is not a common planetary system, and  therefore we do not need a statistical success, but just a possible outcome from our simulations. Thus, with a 1\% success rate, this work shows that the Solar System could have emerged from a low-viscosity protoplanetary disc. 
{To date, we have not detected any planetary system comparable to the four giants of our Solar System, and hence this configuration does not need to be the most frequent outcome of our simulations.} 
How the results depend on the planetesimal disc properties in detail {(e.g. timing of the instability)} is far beyond the scope of this paper and will be the object of a future work.




\begin{acknowledgements}
We thank the reviewer for their valuable comments which improved this manuscript.  This work has been supported by the French government, through the $ \mathrm{UCA^{J.E.D.I.}} $ Investments in the Future project managed by the National Research Agency (ANR) with the reference number ANR-15-IDEX-01. 
This work was granted access to the HPC resources of IDRIS under the allocations A0140407233  and  A0160407233  made by GENCI and from "Mesocentre SIGAMM" hosted by Observatoire de la C\^ote d'Azur.
The authors acknowledge funding from the ERC project N. 101019380 “HolyEarth”.
LE whish to thank Alain Miniussi for maintainance and re-factorization of the code FARGOCA

\end{acknowledgements}
\bibliographystyle{aa}

\bibliography{Article_NM}

\begin{thebibliography}{61}
\expandafter\ifx\csname natexlab\endcsname\relax\def\natexlab#1{#1}\fi

\bibitem[{Agnor \& Lin(2012)}]{agnorMigrationJupiterSaturn2012}
Agnor, C.~B. \& Lin, D. N.~C. 2012, The Astrophysical Journal, 745, 143

\bibitem[{Avdellidou {et~al.}(2024)Avdellidou, Delbo', Nesvorn{\'y}, Walsh, \&
  Morbidelli}]{avdellidouDatingSolarSystem2024}
Avdellidou, C., Delbo', M., Nesvorn{\'y}, D., Walsh, K.~J., \& Morbidelli, A.
  2024, Science, 384, 348

\bibitem[{Batygin {et~al.}(2019)Batygin, Adams, Brown, \&
  Becker}]{batyginPlanetNineHypothesis2019}
Batygin, K., Adams, F.~C., Brown, M.~E., \& Becker, J.~C. 2019, Physics
  Reports, 805, 1

\bibitem[{Batygin \& Brown(2010)}]{batyginEarlyDynamicalEvolution2010}
Batygin, K. \& Brown, M.~E. 2010, The Astrophysical Journal, 716, 1323

\bibitem[{Batygin \& Brown(2016)}]{batyginEvidenceDistantGiant2016}
Batygin, K. \& Brown, M.~E. 2016, The Astronomical Journal, 151, 22

\bibitem[{Batygin \& Petit(2023)}]{batyginDissipativeCapturePlanets2023}
Batygin, K. \& Petit, A.~C. 2023, The Astrophysical Journal, 946, L11

\bibitem[{Bitsch {et~al.}(2020)Bitsch, Trifonov, \&
  Izidoro}]{bitschEccentricityDistributionGiant2020a}
Bitsch, B., Trifonov, T., \& Izidoro, A. 2020, Astronomy \& Astrophysics, 643,
  A66

\bibitem[{Brasser {et~al.}(2009)Brasser, Morbidelli, Gomes, Tsiganis, \&
  Levison}]{brasserConstructingSecularArchitecture2009}
Brasser, R., Morbidelli, A., Gomes, R., Tsiganis, K., \& Levison, H.~F. 2009,
  Astronomy and Astrophysics, 507, 1053

\bibitem[{Brown \& Batygin(2021)}]{brownOrbitPlanetNine2021}
Brown, M.~E. \& Batygin, K. 2021, The Astronomical Journal, 162, 219

\bibitem[{Chiang \& Youdin(2010)}]{chiangFormingPlanetesimalsSolar2010}
Chiang, E. \& Youdin, A.~N. 2010, Annual Review of Earth and Planetary
  Sciences, 38, 493

\bibitem[{Clement {et~al.}(2021{\natexlab{a}})Clement, Deienno, Kaib, Izidoro,
  Raymond, \& Chambers}]{clementBornExtraeccentricBroad2021}
Clement, M.~S., Deienno, R., Kaib, N.~A., {et~al.} 2021{\natexlab{a}}, Icarus,
  367, 114556

\bibitem[{Clement {et~al.}(2019)Clement, Kaib, Raymond, Chambers, \&
  Walsh}]{clementEarlyInstabilityScenario2019}
Clement, M.~S., Kaib, N.~A., Raymond, S.~N., Chambers, J.~E., \& Walsh, K.~J.
  2019, Icarus, 321, 778

\bibitem[{Clement {et~al.}(2018)Clement, Kaib, Raymond, \&
  Walsh}]{clementMarsGrowthStunted2018}
Clement, M.~S., Kaib, N.~A., Raymond, S.~N., \& Walsh, K.~J. 2018, Icarus, 311,
  340

\bibitem[{Clement {et~al.}(2021{\natexlab{b}})Clement, Raymond, Kaib, Deienno,
  Chambers, \& Izidoro}]{clementBornEccentricConstraints2021}
Clement, M.~S., Raymond, S.~N., Kaib, N.~A., {et~al.} 2021{\natexlab{b}},
  Icarus, 355, 114122

\bibitem[{Cresswell \&
  Nelson(2008)}]{cresswellThreedimensionalSimulationsMultiple2008}
Cresswell, P. \& Nelson, R.~P. 2008, Astronomy and Astrophysics, 482, 677

\bibitem[{Crida(2009)}]{cridaMinimumMassSolar2009}
Crida, A. 2009, The Astrophysical Journal, 698, 606

\bibitem[{{De Val-Borro} {et~al.}(2006){De Val-Borro}, Edgar, Artymowicz,
  Ciecielag, Cresswell, D'Angelo, {Delgado-Donate}, Dirksen, Fromang,
  Gawryszczak, Klahr, Kley, Lyra, Masset, Mellema, Nelson, Paardekooper,
  Peplinski, Pierens, Plewa, Rice, Schafer, \&
  Speith}]{deval-borroComparativeStudyDiscplanet2006}
{De Val-Borro}, M., Edgar, R.~G., Artymowicz, P., {et~al.} 2006, Monthly
  Notices of the Royal Astronomical Society, 370, 529

\bibitem[{Deienno {et~al.}(2017)Deienno, Morbidelli, Gomes, \&
  Nesvorn{\'y}}]{deiennoConstrainingGiantPlanets2017}
Deienno, R., Morbidelli, A., Gomes, R.~S., \& Nesvorn{\'y}, D. 2017, The
  Astronomical Journal, 153, 153

\bibitem[{Dr{\k a}{\.z}kowska {et~al.}(2016)Dr{\k a}{\.z}kowska, Alibert, \&
  Moore}]{drazkowskaCloseinPlanetesimalFormation2016}
Dr{\k a}{\.z}kowska, J., Alibert, Y., \& Moore, B. 2016, Astronomy and
  Astrophysics, 594, A105

\bibitem[{Dullemond {et~al.}(2018)Dullemond, Birnstiel, Huang, Kurtovic,
  Andrews, Guzm{\'a}n, P{\'e}rez, Isella, Zhu, Benisty, Wilner, Bai, Carpenter,
  Zhang, \& Ricci}]{dullemondDiskSubstructuresHigh2018}
Dullemond, C.~P., Birnstiel, T., Huang, J., {et~al.} 2018, The Astrophysical
  Journal, 869, L46

\bibitem[{Gomes {et~al.}(2005)Gomes, Levison, Tsiganis, \&
  Morbidelli}]{gomesOriginCataclysmicLate2005}
Gomes, R., Levison, H.~F., Tsiganis, K., \& Morbidelli, A. 2005, Nature, 435,
  466

\bibitem[{Griveaud {et~al.}(2023)Griveaud, Crida, \&
  Lega}]{griveaudMigrationPairsGiant2023}
Griveaud, P., Crida, A., \& Lega, E. 2023, Astronomy \& Astrophysics, 672, A190

\bibitem[{Izidoro {et~al.}(2024)Izidoro, Deienno, Raymond, \&
  Clement}]{izidoroLinkAthorMeteorites2024}
Izidoro, A., Deienno, R., Raymond, S.~N., \& Clement, M.~S. 2024, The Link
  between {{Athor}} and {{EL}} Meteorites Does Not Constrain the Timing of the
  Giant Planet Instability

\bibitem[{Jacobson {et~al.}(2014)Jacobson, Morbidelli, Raymond, O'Brien, Walsh,
  \& Rubie}]{jacobsonHighlySiderophileElements2014}
Jacobson, S.~A., Morbidelli, A., Raymond, S.~N., {et~al.} 2014, Nature, 508, 84

\bibitem[{Kleine {et~al.}(2009)Kleine, Touboul, Bourdon, Nimmo, Mezger, Palme,
  Jacobsen, Yin, \& Halliday}]{kleineHfWChronologyAccretion2009}
Kleine, T., Touboul, M., Bourdon, B., {et~al.} 2009, Geochimica et Cosmochimica
  Acta, 73, 5150

\bibitem[{Lee \& Peale(2002)}]{leeDynamicsOriginOrbital2002}
Lee, M.~H. \& Peale, S.~J. 2002, The Astrophysical Journal, 567, 596

\bibitem[{Lesur {et~al.}(2022)Lesur, Ercolano, Flock, Lin, Yang, Barranco,
  {Benitez-Llambay}, Goodman, Johansen, Klahr, Laibe, Lyra, Marcus, Nelson,
  Squire, Simon, Turner, Umurhan, \&
  Youdin}]{lesurHydroMagnetohydroDustGas2022}
Lesur, G., Ercolano, B., Flock, M., {et~al.} 2022, Hydro-, {{Magnetohydro-}},
  and {{Dust-Gas Dynamics}} of {{Protoplanetary Disks}}

\bibitem[{Levison {et~al.}(2011)Levison, Morbidelli, Tsiganis, Nesvorn{\'y}, \&
  Gomes}]{levisonLATEORBITALINSTABILITIES2011}
Levison, H.~F., Morbidelli, A., Tsiganis, K., Nesvorn{\'y}, D., \& Gomes, R.
  2011, The Astronomical Journal, 142, 152

\bibitem[{Liu {et~al.}(2022)Liu, Raymond, \&
  Jacobson}]{liuEarlySolarSystem2022}
Liu, B., Raymond, S.~N., \& Jacobson, S.~A. 2022, Nature, 604, 643

\bibitem[{Masset \& Snellgrove(2001)}]{massetReversingTypeII2001}
Masset, F. \& Snellgrove, M. 2001, Monthly Notices of the Royal Astronomical
  Society, 320, L55

\bibitem[{Masset {et~al.}(2006)Masset, Morbidelli, Crida, \&
  Ferreira}]{massetDiskSurfaceDensity2006}
Masset, F.~S., Morbidelli, A., Crida, A., \& Ferreira, J. 2006, The
  Astrophysical Journal, 642, 478

\bibitem[{Matsumura {et~al.}(2021)Matsumura, Brasser, \&
  Ida}]{matsumuraNbodySimulationsPlanet2021}
Matsumura, S., Brasser, R., \& Ida, S. 2021, Astronomy and Astrophysics, 650,
  A116

\bibitem[{Minton \& Malhotra(2011)}]{mintonSecularResonanceSweeping2011}
Minton, D.~A. \& Malhotra, R. 2011, The Astrophysical Journal, 732, 53

\bibitem[{Morbidelli {et~al.}(2022)Morbidelli, Bailli{\'e}, Batygin, Charnoz,
  Guillot, Rubie, \& Kleine}]{morbidelliContemporaryFormationEarly2022}
Morbidelli, A., Bailli{\'e}, K., Batygin, K., {et~al.} 2022, Nature Astronomy,
  6, 72

\bibitem[{Morbidelli {et~al.}(2010)Morbidelli, Brasser, Gomes, Levison, \&
  Tsiganis}]{morbidelliEvidenceAsteroidBelt2010}
Morbidelli, A., Brasser, R., Gomes, R., Levison, H.~F., \& Tsiganis, K. 2010,
  The Astronomical Journal, 140, 1391

\bibitem[{Morbidelli \& Crida(2007)}]{morbidelliDynamicsJupiterSaturn2007}
Morbidelli, A. \& Crida, A. 2007, Icarus, 191, 158

\bibitem[{Morbidelli {et~al.}(2005)Morbidelli, Levison, Tsiganis, \&
  Gomes}]{morbidelliChaoticCaptureJupiter2005}
Morbidelli, A., Levison, H.~F., Tsiganis, K., \& Gomes, R. 2005, Nature, 435,
  462

\bibitem[{Morbidelli {et~al.}(2018)Morbidelli, Nesvorny, Laurenz, Marchi,
  Rubie, {Elkins-Tanton}, Wieczorek, \&
  Jacobson}]{morbidelliTimelineLunarBombardment2018}
Morbidelli, A., Nesvorny, D., Laurenz, V., {et~al.} 2018, Icarus, 305, 262

\bibitem[{Morbidelli {et~al.}(2007)Morbidelli, Tsiganis, Crida, Levison, \&
  Gomes}]{morbidelliDynamicsGiantPlanets2007}
Morbidelli, A., Tsiganis, K., Crida, A., Levison, H.~F., \& Gomes, R. 2007, The
  Astronomical Journal, 134, 1790

\bibitem[{Nesvorn{\'y}(2018)}]{nesvornyDynamicalEvolutionEarly2018}
Nesvorn{\'y}, D. 2018, Annual Review of Astronomy and Astrophysics, 56, 137

\bibitem[{Nesvorn{\'y} \& Morbidelli(2012)}]{nesvornySTATISTICALSTUDYEARLY2012}
Nesvorn{\'y}, D. \& Morbidelli, A. 2012, The Astronomical Journal, 144, 117

\bibitem[{Nesvorn{\'y} {et~al.}(2021)Nesvorn{\'y}, Roig, \&
  Deienno}]{nesvornyRoleEarlyGiantplanet2021}
Nesvorn{\'y}, D., Roig, F.~V., \& Deienno, R. 2021, The Astronomical Journal,
  161, 50

\bibitem[{Nesvorn{\'y} {et~al.}(2023)Nesvorn{\'y}, Roig, Vokrouhlick{\'y},
  Bottke, Marchi, Morbidelli, \& Deienno}]{nesvornyEarlyBombardmentMoon2023}
Nesvorn{\'y}, D., Roig, F.~V., Vokrouhlick{\'y}, D., {et~al.} 2023, Icarus,
  399, 115545

\bibitem[{Pichierri {et~al.}(2024)Pichierri, Bitsch, \&
  Lega}]{pichierriRecipeEccentricityInclination2024}
Pichierri, G., Bitsch, B., \& Lega, E. 2024, A Recipe for Eccentricity and
  Inclination Damping for Partial Gap Opening Planets in {{3D}} Disks

\bibitem[{Pierens {et~al.}(2014)Pierens, Raymond, Nesvorny, \&
  Morbidelli}]{pierensOutwardMigrationJupiter2014}
Pierens, A., Raymond, S.~N., Nesvorny, D., \& Morbidelli, A. 2014, The
  Astrophysical Journal, 795, L11

\bibitem[{Pinte {et~al.}(2016)Pinte, Dent, M{\'e}nard, Hales, Hill, Cortes, \&
  {de Gregorio-Monsalvo}}]{pinteDustGasDisk2016}
Pinte, C., Dent, W. R.~F., M{\'e}nard, F., {et~al.} 2016, The Astrophysical
  Journal, 816, 25

\bibitem[{Rein {et~al.}(2019)Rein, Hernandez, Tamayo, Brown, Eckels, Holmes,
  Lau, Leblanc, \& Silburt}]{reinHybridSymplecticIntegrators2019}
Rein, H., Hernandez, D.~M., Tamayo, D., {et~al.} 2019, Monthly Notices of the
  Royal Astronomical Society, 485, 5490

\bibitem[{Rein \& Liu(2012)}]{reinREBOUNDOpensourceMultipurpose2012}
Rein, H. \& Liu, S.~F. 2012, Astronomy and Astrophysics, 537, A128

\bibitem[{Ryder(2000)}]{ryderCatastrophicEventsMass2000}
Ryder, G. 2000, Meteoritics and Planetary Science, 35, 1126

\bibitem[{Savvidou {et~al.}(2020)Savvidou, Bitsch, \&
  Lambrechts}]{savvidouInfluenceGrainGrowth2020}
Savvidou, S., Bitsch, B., \& Lambrechts, M. 2020, Astronomy and Astrophysics,
  640, A63

\bibitem[{Sellek {et~al.}(2020)Sellek, Booth, \&
  Clarke}]{sellekDustyOriginCorrelation2020}
Sellek, A.~D., Booth, R.~A., \& Clarke, C.~J. 2020, Monthly Notices of the
  Royal Astronomical Society, 498, 2845

\bibitem[{Shakura \& Sunyaev(1973)}]{shakuraBlackHolesBinary1973}
Shakura, N.~I. \& Sunyaev, R.~A. 1973, Astronomy and Astrophysics, 24, 337

\bibitem[{Stewart \&
  Wetherill(1988)}]{stewartEvolutionPlanetesimalVelocities1988}
Stewart, G.~R. \& Wetherill, G.~W. 1988, Icarus, 74, 542

\bibitem[{Tera {et~al.}(1974)Tera, Papanastassiou, \&
  Wasserburg}]{teraIsotopicEvidenceTerminal1974}
Tera, F., Papanastassiou, D.~A., \& Wasserburg, G.~J. 1974, Earth and Planetary
  Science Letters, 22, 1

\bibitem[{Thommes {et~al.}(2008)Thommes, Bryden, Wu, \&
  Rasio}]{thommesMeanMotionResonances2008}
Thommes, E.~W., Bryden, G., Wu, Y., \& Rasio, F.~A. 2008, The Astrophysical
  Journal, 675, 1538

\bibitem[{Tsiganis {et~al.}(2005)Tsiganis, Gomes, Morbidelli, \&
  Levison}]{tsiganisOriginOrbitalArchitecture2005}
Tsiganis, K., Gomes, R., Morbidelli, A., \& Levison, H.~F. 2005, Nature, 435,
  459

\bibitem[{Turner {et~al.}(2014)Turner, Fromang, Gammie, Klahr, Lesur, Wardle,
  \& Bai}]{turnerTransportAccretionPlanetForming2014}
Turner, N.~J., Fromang, S., Gammie, C., {et~al.} 2014, Transport and
  {{Accretion}} in {{Planet-Forming Disks}} (eprint: arXiv:1401.7306), 411

\bibitem[{Villenave {et~al.}(2022)Villenave, Stapelfeldt, Duch{\^e}ne,
  M{\'e}nard, Lambrechts, Sierra, Flores, Dent, Wolff, Ribas, Benisty, Cuello,
  \& Pinte}]{villenaveHighlySettledDisk2022}
Villenave, M., Stapelfeldt, K.~R., Duch{\^e}ne, G., {et~al.} 2022, The
  Astrophysical Journal, 930, 11

\bibitem[{Walsh {et~al.}(2011)Walsh, Morbidelli, Raymond, O'Brien, \&
  Mandell}]{walshLowMassMars2011}
Walsh, K.~J., Morbidelli, A., Raymond, S.~N., O'Brien, D.~P., \& Mandell, A.~M.
  2011, Nature, 475, 206

\bibitem[{Zellner(2017)}]{zellnerCataclysmNoMore2017}
Zellner, N. E.~B. 2017, Origins of Life and Evolution of the Biosphere, 47, 261

\bibitem[{Zhu {et~al.}(2019)Zhu, Artemieva, Morbidelli, Yin, Becker, \&
  W{\"u}nnemann}]{zhuReconstructingLateaccretionHistory2019}
Zhu, M.-H., Artemieva, N., Morbidelli, A., {et~al.} 2019, Nature, 571, 226

\end{thebibliography}

\clearpage

\end{document}